\documentclass[twocolumn,journal]{IEEEtran}
\usepackage[T1]{fontenc}
\usepackage[latin9]{inputenc}
\usepackage{color}
\usepackage{array}
\usepackage{units}
\usepackage{bm}
\usepackage{multirow}
\usepackage{amsmath}
\usepackage{amsthm}
\usepackage{amssymb}
\usepackage{graphicx}
\usepackage[unicode=true,
 bookmarks=true,bookmarksnumbered=true,bookmarksopen=true,bookmarksopenlevel=1,
 breaklinks=false,pdfborder={0 0 0},pdfborderstyle={},backref=false,colorlinks=false]
 {hyperref}
\hypersetup{pdftitle={Your Title},
 pdfauthor={Your Name},
 pdfpagelayout=OneColumn, pdfnewwindow=true, pdfstartview=XYZ, plainpages=false}

\makeatletter

\providecommand{\tabularnewline}{\\}

\theoremstyle{plain}
\newtheorem{thm}{\protect\theoremname}
\theoremstyle{remark}
\newtheorem{rem}[thm]{\protect\remarkname}

\usepackage[caption=false,font=footnotesize]{subfig}
\usepackage{diagbox}
\usepackage{algorithm}
\usepackage{algorithmic}
\usepackage{color}
\usepackage{amsmath}
\usepackage{bm}

\@ifundefined{showcaptionsetup}{}{%
 \PassOptionsToPackage{caption=false}{subfig}}
\usepackage{subfig}
\makeatother

\providecommand{\remarkname}{Remark}
\providecommand{\theoremname}{Theorem}

\begin{document}
\title{Bayes-Optimal Unsupervised Learning for Channel Estimation in Near-Field
Holographic MIMO}
\author{Wentao~Yu,~\IEEEmembership{Graduate Student Member,~IEEE,} Hengtao~He,~\IEEEmembership{Member,~IEEE,}
Xianghao~Yu,~\\\IEEEmembership{Senior Member,~IEEE,} Shenghui
Song,~\IEEEmembership{Senior Member,~IEEE,} Jun~Zhang,~\IEEEmembership{Fellow,~IEEE,}
\\Ross~Murch,~\IEEEmembership{Fellow,~IEEE,}~and~Khaled~B.
Letaief,~\IEEEmembership{Fellow,~IEEE}\thanks{Manuscript received 15 December 2023; revised 22 April 2024; accepted
5 June 2024. This work was supported in part by the Hong Kong Research
Grants Council under Grant No. 21215423, 16209622 and the Areas of
Excellence Scheme (Grant No. AoE/E-601/22-R), and in part by the NSFC/RGC
Joint Research Scheme sponsored by the Research Grants Council of
the Hong Kong Special Administrative Region, China and National Natural
Science Foundation of China (Project No. N\_HKUST656/22). An earlier
version of this paper was presented in part at the IEEE International
Conference on Communications (ICC), Denver, CO, USA, Jun. 2024 \cite{2024Yu}.
The guest editor coordinating the review of this manuscript and approving
it for publication was Prof. Özlem Tu\u{g}fe Demir. \textit{(Corresponding
author: Hengtao He.)}}\thanks{Wentao Yu, Hengtao He, Shenghui Song, Jun Zhang, Ross Murch, and Khaled
B. Letaief are with the Department of Electronic and Computer Engineering,
The Hong Kong University of Science and Technology (HKUST), Kowloon,
Hong Kong (e-mail: wyuaq@connect.ust.hk; eehthe@ust.hk; eeshsong@ust.hk;
eejzhang@ust.hk; eermurch@ust.hk; eekhaled@ust.hk).}\thanks{Xianghao Yu is with the Department of Electrical Engineering, City
University of Hong Kong, Kowloon, Hong Kong (e-mail: alex.yu@cityu.edu.hk).}\thanks{The source code is publicly available at \protect\href{https://github.com/wyuaq}{https://github.com/wyuaq}. }}
\markboth{Accepted by IEEE Journal of Selected Topics in Signal Processing}{}
\maketitle
\begin{abstract}
Holographic MIMO (HMIMO) is being increasingly recognized as a key
enabling technology for 6G wireless systems through the deployment
of an extremely large number of antennas within a compact space to
fully exploit the potentials of the electromagnetic (EM) channel.
Nevertheless, the benefits of HMIMO systems cannot be fully unleashed
without an efficient means to estimate the high-dimensional channel,
whose distribution becomes increasingly complicated due to the accessibility
of the near-field region. In this paper, we address the fundamental
challenge of designing a \textit{low-complexity Bayes-optimal} channel
estimator in near-field HMIMO systems operating in \textit{unknown}
EM environments. The core idea is to estimate the HMIMO channels solely
based on the Stein\textquoteright s score function of the received
pilot signals and an estimated noise level, \textit{without} relying
on priors or supervision that is not feasible in practical deployment.
A neural network is trained with the unsupervised denoising score
matching objective to learn the parameterized score function. Meanwhile,
a principal component analysis (PCA)-based algorithm is proposed to
estimate the noise level leveraging the low-rank near-field spatial
correlation. Building upon these techniques, we develop a Bayes-optimal
score-based channel estimator for fully-digital HMIMO transceivers
in a closed form. The optimal score-based estimator is also extended
to hybrid analog-digital HMIMO systems by incorporating it into a
low-complexity message passing algorithm. The (quasi-) Bayes-optimality
of the proposed estimators is validated both in theory and by extensive
simulation results. In addition to optimality, it is shown that our
proposal is robust to various mismatches and can quickly adapt to
dynamic EM environments in an online manner thanks to its unsupervised
nature, demonstrating its potential in real-world deployment. 
\end{abstract}


\section{Introduction\label{sec:Introduction}}

With an ultra-massive number of antennas closely packed in a compact
space, holographic MIMO (HMIMO) is envisioned as a promising next-generation
multi-antenna technology to enable an unprecedentedly high degree
of spectral and energy efficiency \cite{2020Pizzo}. Extensive research
efforts have been dedicated to many different aspects of HMIMO systems,
such\textcolor{black}{{} as fundamental performance limits \cite{2024Zhang-foundamental}},
channel measurement and modeling \cite{2020Pizzo,2022Pizzo,2022Demir,2022Wei,2023Wang-measurement},
beamforming and focusing methods \cite{2022Deng,2023An-JSAC,2023Wei,2023Zhang-analog},
positioning and sensing \cite{2022Damico,2023Elzanaty}, and integrated
sensing and communications \cite{2022Zhang}, etc. Nevertheless, one
fundamental problem is that, the promised gains of HMIMO can hardly
be realized without an efficient means to estimate the extremely high-dimensional
channel. 

The challenges to estimate the HMIMO channel are mainly three-fold
\cite{2023Yu-AI}. The first challenge arises from the complicated
and dynamic channel distributions. The extremely large-scale array
significantly expands the near-field region with spherical wavefront,
which greatly complicates the channel distributions and enlarges the
codebook size to sparsify the channel \cite{2022Cui}. Also, since
HMIMO systems often operate at higher frequency bands, they are vulnerable
to various types of blockage and movement, which can easily shift
the channel distributions in a dynamic manner. Second, priors or a
supervised dataset are difficult to obtain. Existing estimators either
utilized the prior knowledge, e.g., sparsity in some transform domain,
or a large supervised channel dataset, to enhance the estimation accuracy.
Nevertheless, both of them require tedious, if not prohibitive, efforts
to realize in practice, considering the enormous system scale and
the complicated channel distributions in the near-field HMIMO systems.
The computational complexity poses the third challenge. The scale
of near-field HMIMO systems prohibits the use of classical Bayes-optimal
channel estimators which include complicated matrix inversion, and
calls for low-complexity alternatives. 

\subsection{Related Works}

The simplest approach to estimating the channel is the least squares
(LS) method \cite{2014Lin}, which offers reasonable performance in
fully-digital systems but suffers from a high pilot overhead in hybrid
analog-digital systems. To enhance the performance, the regularized
LS estimators prevail, which, in addition to the LS objective, minimize
an extra regularization term to enforce the channel to follow the
prior knowledge of its characteristics, e.g., sparsity \cite{2023Liu,2024Zhang-Sparse,2024Cao}
or low-rankness \cite{2016Xie}. By contrast, the Bayesian point of
view instead believes that the channels should follow a prior distribution,
and focuses on optimizing the maximum a posteriori (MAP) or the minimum
mean-square-error (MMSE) objectives \cite{2021Zhu}. Both the regularization-based
and Bayesian perspectives stress the availability of a proper prior.
Nevertheless, finding such a prior may be difficult in practice as
the HMIMO channels have become increasingly complicated, e.g., with
the near-field \cite{2022Cui} and hybrid-field effects \cite{2023Yu-JSTSP}.
Simply resorting to the classical angular domain sparsity or the priors
based on some simple mixture distributions can no longer offer competitive
accuracy. 

Deep learning (DL)-based approaches then emerge as a potential paradigm
that attempts to learn the priors from a large volume of data instead
of relying on hand-crafted designs \cite{2019He,2023Zhang-Model}.
As a result, data is fundamental to DL-based methods. Depending on
how the data are leveraged, DL methods can be categorized as supervised,
unsupervised, and generative approaches. As for the channel estimation
problem, supervised learning refers to those that require a paired
dataset of the received pilot signals $\mathbf{y}$ and the channel
$\mathbf{h}$, i.e., $(\mathbf{y},\mathbf{h})$, in order to implicitly
exploit the prior, while the generative learning requires a dataset
of $\mathbf{h}$ to explicitly model the prior distribution \cite{2021Balevi}.
However, both supervised and generative learning requires a dataset
involving the ground-truth channel $\mathbf{h}$, which, in practice,
cannot be easily acquired without a competitive channel estimator.
This brings about a chicken-and-egg dilemma: To train competitive
models we need to prepare a dataset of ground-truth channels, but
such a channel dataset comes from nowhere. To tackle the dilemma,
unsupervised learning is an ideal candidate as it only requires a
dataset of the received pilots $\mathbf{y}$, which can be readily
obtained in abundant volume. Existing unsupervised channel estimators
generally suffer from sub-optimal accuracy. In \cite{2021Zheng},
the authors proposed to train a multi-layer perceptron (MLP)-based
neural network to minimize online loss functions based on the LS and
the nuclear-norm-based regularizer, respectively. This is, essentially,
training a neural network-based solver for the regularization-based
channel estimator, whose performance is hence still subject to the
hand-crafted regularizer. Results therein show a large performance
gap compared to supervised learning. In \cite{2023He,2023Yu}, the
authors proposed to minimize the Stein's unbiased risk estimator (SURE)
objective as a surrogate to the mean-square-error (MSE) loss in order
to approach the accuracy of supervised learning, but the SURE loss
involves a Monte-Carlo approximation, which will degrade the accuracy
of the trained estimator and increase its bias. The performance is
still inferior to supervised training, let alone the oracle performance
bound. 

As a golden standard, the MMSE estimator can achieve the Bayes-optimal
accuracy in terms of MSE, but as mentioned in the above discussions,
its implementation requires either a perfect knowledge of the prior
distribution \cite{2022Demir,2023An}, or learning such a distribution
from a substantial number of ground-truth channels \cite{2023Yu-JSTSP,2023Yu-AI},
both of which are difficult, if not impossible, to realize in near-field
HMIMO systems. In addition, computational complexity is again a significant
issue. Even the linear MMSE (LMMSE) estimator still contains a general
matrix inverse that consumes a significant amount of computational
budget \cite{2020He}. The existing works on HMIMO channel estimation
mainly focused on the low-complexity alternatives to the MMSE estimator.
In \cite{2022Demir}, a subspace-based channel estimation algorithm
was proposed, in which the low-rank property of the HMIMO spatial
correlation was exploited without full knowledge of the spatial correlation.
In \cite{2023Damico}, a discrete Fourier transform (DFT)-based HMIMO
channel estimator was proposed based on a circulant matrix-based approximation
of the spatial correlation matrix. Nevertheless, such an algorithm
was limited to the uniform linear array (ULA)-based HMIMO systems,
and cannot be extended to more general antenna array geometries. In
\cite{2021Wan}, angular domain channel sparsity was exploited for
the design of compressed sensing (CS)-based channel estimators in
holographic reconfigurable intelligent surface (RIS)-aided wireless
systems, but such a nice property dose not hold in the considered
near-field region. In \cite{2023An}, a concise tutorial of HMIMO
channel modeling and estimation was presented. Even though the aforementioned
estimators significantly outperform the conventional LS scheme, there
still exists quite a large gap from that of the Bayes-optimal MMSE
estimator. 

\subsection{Contributions}

The target of this paper is thus to establish an efficient and Bayes-optimal
channel estimator for near-field HMIMO systems that operates in an
arbitrary unknown EM environment, circumventing t\textcolor{black}{he
need for p}riors or supervision. The main contributions are summarized
as follows. 
\begin{itemize}
\item We introduce an \textit{unsupervised} DL framework that validates
the feasibility of establishing Bayes-optimal channel estimators solely
based on the \textit{Stein's score function} of the received pilot
signals and the estimated noise level. 
\item We then devise practical algorithms to obtain the two key ingredients
of the unsupervised Bayes-optimal estimator, i.e., the score function
and the noise level. For the former, we propose to train a neural
network-based parameterized score function with the denoising score
matching loss. As for the latter, we exploit the low-rankness of the
near-field spatial correlation to build a principal component analysis
(PCA)-based noise level estimator. The theoretical underpinnings are
discussed along with the algorithms. 
\item We propose score-based channel estimators for near-field HMIMO systems
with both the fully-digital and hybrid analog-digital architectures.
For fully-digital transceivers, the Bayes-optimal estimator is as
simple as a sum of the received pilot signals and the rescaled score
function by the estimated noise level. As for the hybrid analog-digital
architecture, we insert the proposed score-based estimator into iterative
message passing algorithms, and propose an efficient technique to
reduce their complexity. A complexity analysis shows that the algorithms
in both setups are only dominated by the matrix-vector product. 
\item Extensive numerical results demonstrate that the proposed score-based
estimator is indeed quasi-optimal when compared to the oracle performance
bound. In addition, the robustness and the generalization capability
are validated in different system configurations. Furthermore, thanks
to the unsupervised nature, the proposed estimator can automatically
adapt to dynamic environments with varying channel distributions,
which justifies its practical value. 
\end{itemize}

\subsection{Paper Organization and Notation}

The remaining parts of the article are organized as follows. In Section
\ref{sec:2-HMIMO-Channel-and}, we introduce the statistical channel
model and the system model for near-field HMIMO systems. In Section
\ref{sec:Bayes-Optimal-Unsupervised-Chann}, we first discuss the
general framework and the two core techniques of the score-based estimator,
i.e., the unsupervised learning of the score function and the PCA-based
noise level estimation. Then, in Section \ref{sec:Practical-Algorithms-of},
we respectively present the algorithms for the fully-digital, as well
as the hybrid analog-digital systems. In Section \ref{sec:Simulation-Results},
extensive simulation results are presented to illustrate the advantages
of our proposal in terms of performance, generalization, and robustness.
In Section \ref{sec:Conclusion-and-Future}, we conclude the paper
and discuss future directions. 

\textit{Notation:} $a$ is a scalar. $\text{tr}(\mathbf{a})$, $\|\mathbf{a}\|$,
and $\mathbf{a}(n)$ are the trace, the $\ell_{2}$-norm, and the
$n$-th entry of a vector $\mathbf{a}$, respectively. $\mathbf{A}^{T}$,
$\mathbf{A}^{H}$, $\text{\ensuremath{\Re}}(\mathbf{A})$, $\ensuremath{\Im}(\mathbf{A})$,
$\mathbf{A}(n,m)$ are the transpose, Hermitian, real part, imaginary
part, and the $(n,m)$-th entry of a matrix $\mathbf{A}$, respectively.
$\mathcal{CN}(\boldsymbol{\mu},\mathbf{R})$ and $\mathcal{N}(\boldsymbol{\mu},\mathbf{R})$
are the complex and the real Gaussian distributions with mean $\boldsymbol{\mu}$
and covariance $\mathbf{R}$, respectively. $\mathbf{I}$ is an identity
matrix with appropriate size. 

\section{HMIMO Channel and System Models\label{sec:2-HMIMO-Channel-and}}

\subsection{Channel Model for Near-Field HMIMO}

\begin{figure}[t]
\centering{}\textcolor{black}{\includegraphics[width=6.5cm]{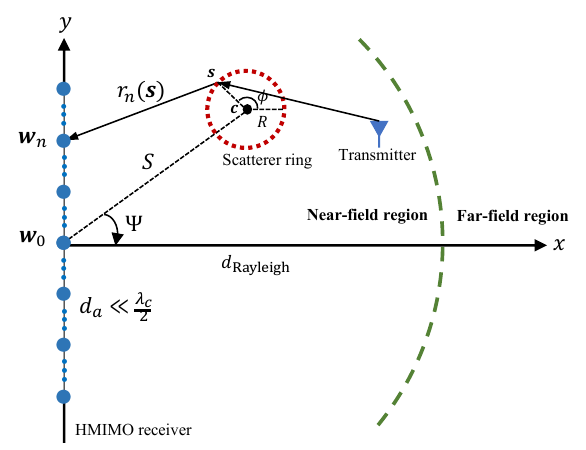}\caption{A ULA-shaped HMIMO BS in the cartesian coordinate system, where the
transmitter and the scatterers locate in the near field of the HMIMO
array. \label{fig:system-model}}
}
\end{figure}

Consider the uplink of an HMIMO system where the base station (BS)
is equipped with a uniform linear array (ULA)\footnote{\textcolor{black}{The applicability of the proposed score-based algorithms
extends beyond the ULA model and encompasses various array configurations.
As one will see from Section \ref{sec:Bayes-Optimal-Unsupervised-Chann},
the derivation of the algorithms rely solely on the low-rank property
of the near-field spatial correlation matrix, rather than the specific
array geometries. The ULA-based channel model was adopted because,
at the time of preparing this paper, we could only find one reference
in the literature, namely \cite{2022Dong}, on the near-field spatial
correlation for HMIMO systems. }} with thousands of compactly packed antennas with spacings $d_{a}$
being much smaller than half of the carrier wavelength $\lambda_{c}$,
i.e., $d_{a}\ll\frac{\lambda_{c}}{2}$, as illustrated in Fig. \ref{fig:system-model}.
The BS is assumed to simultaneously serve multiple single-antenna
users. We define a Cartesian coordinate system at the center of the
ULA (i.e., the reference antenna), denoted by $\mathbf{w}_{0}$. Similarly,
we denote the position of the $n$-th antenna\textcolor{black}{{} as
$\mathbf{w}_{n}$. The propagation characteristics of the channel
differ in the (radiating) near-field and far-field regions, whose
boundary is determined by the Rayleigh distance $d_{\text{Rayleigh}}\triangleq\frac{2[(N-1)d_{a}]^{2}}{\lambda_{c}}$,
where $N$ refers to the number of antennas. Due to the joint effect
of the array aperture and the carrier wavelength, the near-field region
is non-negligible in HMIMO systems, which, compared to the traditional
far-field setup, is a paradigm shift that brings new degrees-of-freedom
(DoFs) for communications \cite{2022Damico,2024Zhang-Sparse}. }

When there are infinitely many multi-paths, it follows from the central
limit theorem that the channel between the HMIMO and a particular
user, i.e., $\mathbf{\bar{h}}\in\mathbb{C}^{N\times1}$, can be modeled
by the correlated Rayleigh fading\footnote{\textcolor{black}{Similar to \cite{2022Demir,2023An,2023Damico},
in this work, we have also adopted the correlated Rayleigh fading
model. In addition to its mathematical tractability, this choice of
channel model enables the straightforward derivation of the oracle
MMSE performance bound, which serves as an important benchmark for
assessing whether the proposed algorithms have achieved Bayes-optimal
performance.}} as \cite{2022Demir,2023An,2021Bjornson}
\begin{equation}
\mathbf{\bar{h}}\sim\mathcal{CN}(\mathbf{0},\mathbf{R}_{\text{NF}}),\label{eq:NF-channel}
\end{equation}
which is solely determined by the spatial correlation $\mathbf{R}_{\text{NF}}\in\mathbb{C}^{N\times N}$,
with subscript NF denoting the near-field. This model has been and
still is the foundation of most theoretical research in HMIMO systems
\cite{2020Pizzo,2022Demir}. It has been further established in \cite{2022Dong}
that, while the number of scatterers tends to infinity, the near-field
spatial correlation for HMIMO has a generic integral expression, given
by 
\begin{equation}
\mathbf{R}_{\text{NF}}(n,m)=\beta_{0}\int_{\mathbf{s}\in\mathbf{S}}\frac{r_{0}^{2}(\mathbf{s})e^{-j\frac{2\pi}{\lambda_{c}}\left(r_{n}(\mathbf{s})-r_{m}(\mathbf{s})\right)}}{r_{n}(\mathbf{s})r_{m}(\mathbf{s})}f(\mathbf{s})\mathrm{d}\mathbf{s},\label{eq:NF-correlation-integral}
\end{equation}
where $\beta_{0}$ denotes the average received power at the reference
antenna, $\mathbf{s}\in\mathbf{S}$ denotes the position of a scatterer
in the support of random scatterers $\mathbf{S}$, $r_{n}(\mathbf{s})\triangleq\|\mathbf{s}-\mathbf{w}_{n}\|$
refers to the distance between scatterer $\mathbf{s}$ and the $n$-th
antenna element, and $f(\mathbf{s})$ is the probability density function
(PDF) of the scatterer location, called the \textit{power location
spectrum (PLS)}. The expression (\ref{eq:NF-correlation-integral})
holds true for any PLS $f(\mathbf{s})$. 

To gain further insights, we consider a specific $f(\mathbf{s})$,
named the generalized one-ring model, assuming that the scatterers
are located on a ring whose center is freely positioned, as shown
in Fig. 1. We denote the radius and the angular direction of the ring
as $R\geq0$ and $\Psi\in(-\frac{\pi}{2},\frac{\pi}{2})$, respectively.
The coordinate of the ring center is given by $\mathbf{c}=[S\cos\Psi,S\sin\Psi]^{T}$,
where $S\geq0$ is the distance between the origin and the ring center.
The location of a certain scatterer on the ring is then given by $\mathbf{s}=\mathbf{c}+[R\cos\phi,R\sin\phi]^{T}$,
where $\phi\in[-\pi,\pi)$ is the angular position of the scatterer
on the ring. We notice that in a certain EM environment in which $S$,
$R$, and $\Psi$ are given, the distance $r_{n}(\mathbf{s})$ in
(\ref{eq:NF-correlation-integral}) reduces to a function of $\phi$,
given by 
\begin{equation}
r_{n}(\phi)=\sqrt{(S\sin\Psi+R\sin\phi-nd_{a})^{2}+(S\cos\Psi+R\cos\phi)^{2}},\label{eq:distance_r_n}
\end{equation}
so as $f(\mathbf{s})$, which reduces to $f(\phi)$. Plugging (\ref{eq:distance_r_n})
into (\ref{eq:NF-correlation-integral}), when $S\gg R$ is satisfied,
the near-field spatial correlation $\mathbf{R}_{\text{NF}}$ can be
approximately simplified as
\begin{equation}
\begin{aligned}\mathbf{R} & _{\text{NF}}(n,m)\\
 & \thickapprox\beta_{0}\int_{-\pi}^{\pi}\frac{e^{-j\frac{2\pi}{\lambda_{c}}\left(R\left(\frac{b_{n}}{\sqrt{a_{n}}}-\frac{b_{m}}{\sqrt{a_{m}}}\right)+S\left(\sqrt{a_{n}}-\sqrt{a_{m}}\right)\right)}}{\sqrt{a_{n}a_{m}}}f(\phi)\mathrm{d}\phi,
\end{aligned}
\label{eq:NF-correlation-approximation}
\end{equation}
according to \cite[Lemma 2]{2022Dong}. Here $a_{n}$ and $b_{n}$
are respectively defined as $a_{n}\triangleq(nd_{a}/S)^{2}-2(nd_{a}/S)+1$
and $b_{n}\triangleq\cos(\Psi-\phi)-(nd_{a}/S)\sin\phi$, while $a_{m}$
and $b_{m}$ are determined similarly. For the ease of simulations,
we utilize the particular von-Mises distribution for $f(\phi)$ to
model the scatterer distribution on the ring, given by \cite{2002Abdi}
\begin{equation}
f(\phi)=(2\pi I_{0}(\kappa))^{-1}e^{\kappa\cos(\phi-\mu)},\label{eq:von-Mises-distribution}
\end{equation}
in which $I_{0}(\cdot)$ denotes the zeroth-order Bessel function
of the first kind, while $\kappa\geq0$ and $\mu\in[-\pi,\pi)$ are
the two parameters determining the concentration and the mean of the
distribution, respectively. The distribution is the circular analog
of the normal distribution. When $\kappa$ is zero, the distribution
is uniform, while when $\kappa$ is large, the distribution becomes
concentrated about the angle $\mu$. After plugging (\ref{eq:von-Mises-distribution})
into (\ref{eq:NF-correlation-approximation}), the integration can
be computed numerically by Monte-Carlo methods to obtain the near-field
spatial correlation $\mathbf{R}_{\text{NF}}$. 
\begin{figure}[t]
\begin{centering}
\subfloat[]{\centering{}\includegraphics[width=0.23\textwidth]{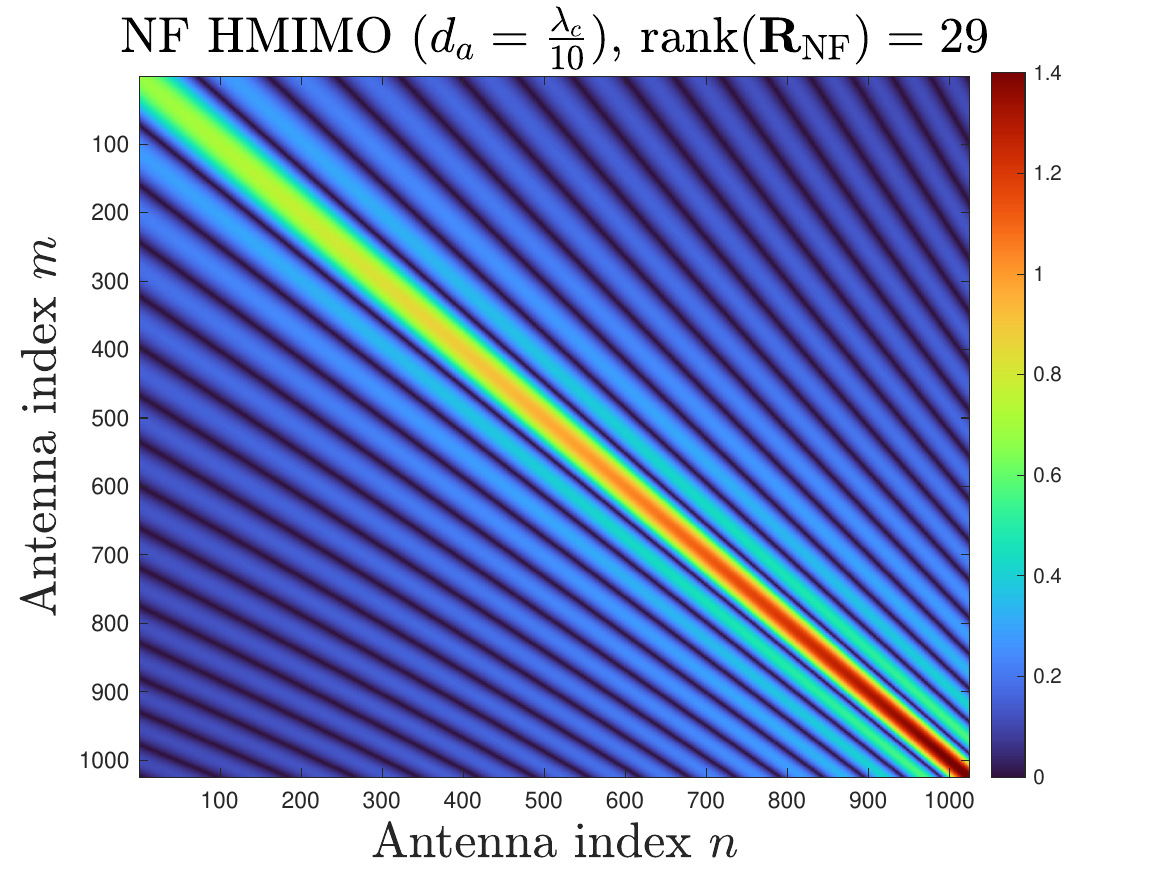}}\subfloat[]{\centering{}\includegraphics[width=0.23\textwidth]{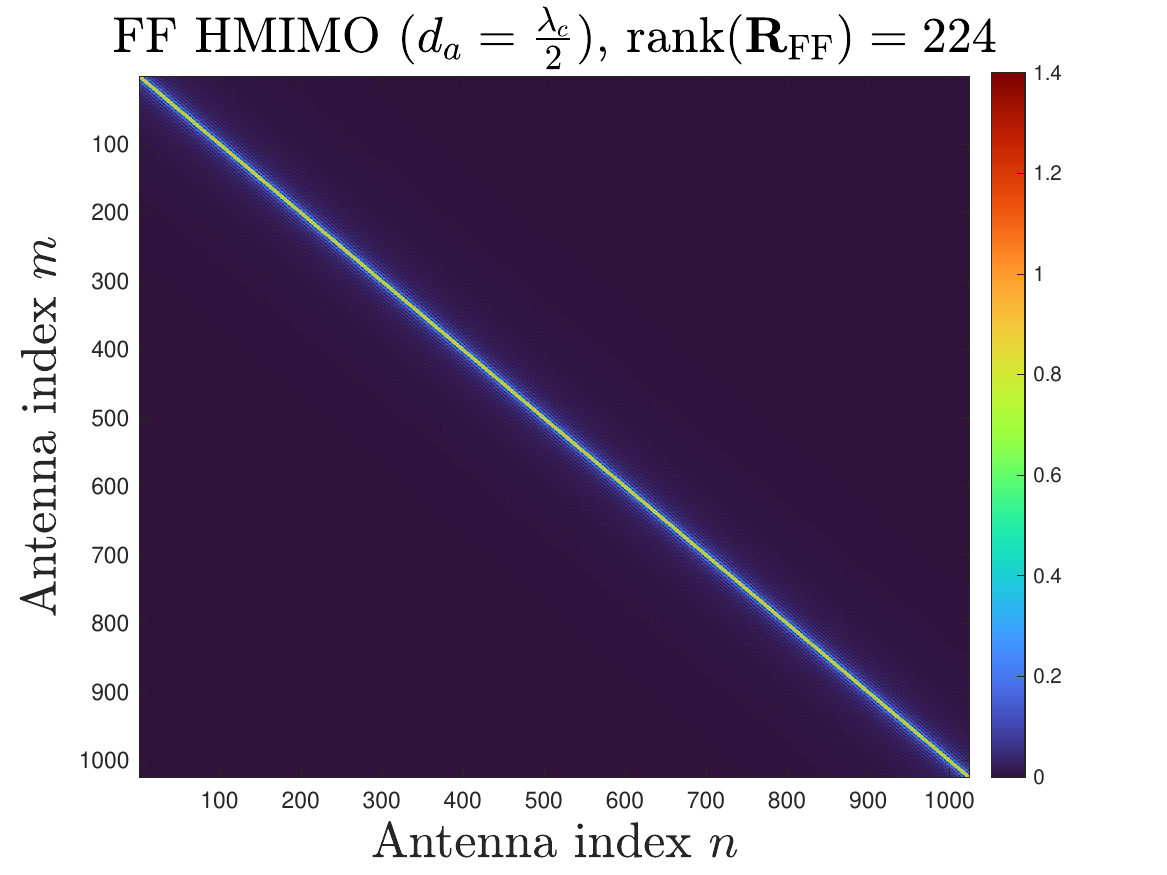}}
\par\end{centering}
\centering{}\subfloat[]{\centering{}\includegraphics[width=0.23\textwidth]{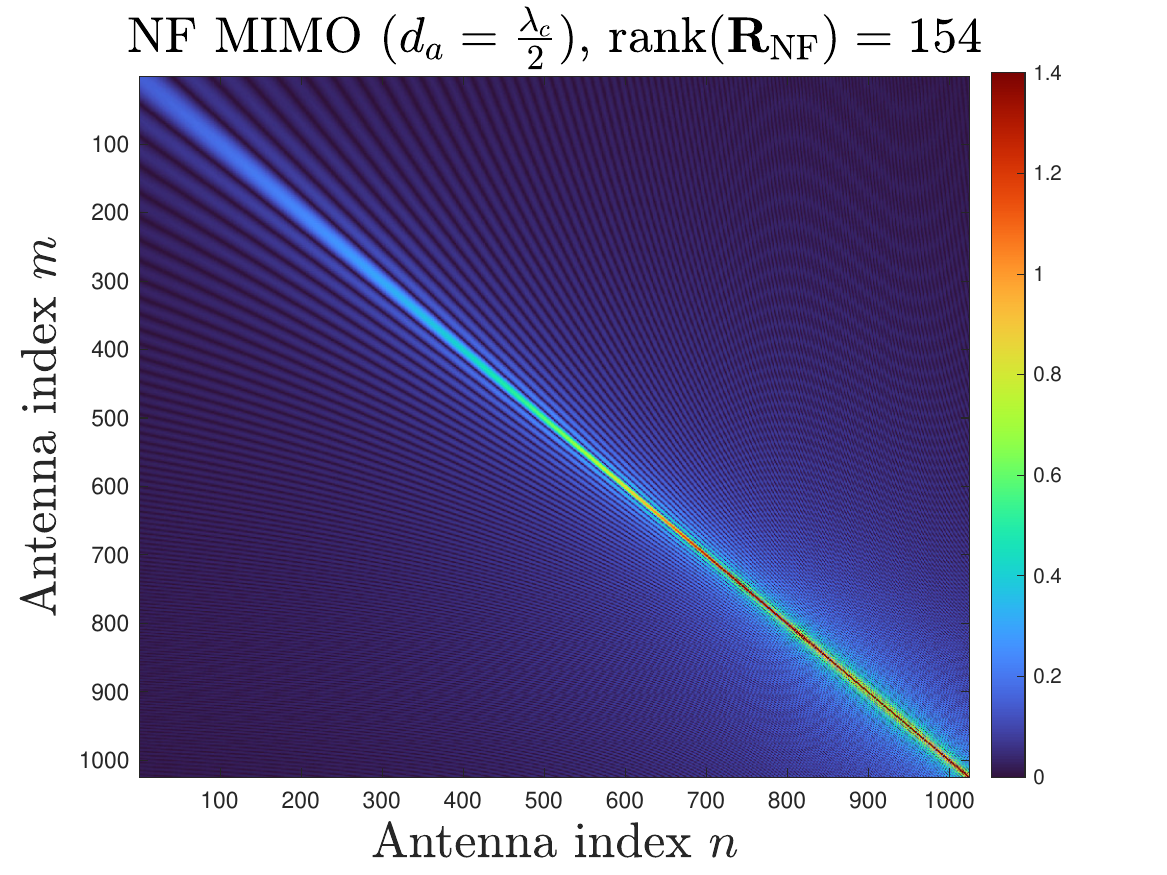}}\subfloat[]{\centering{}\includegraphics[width=0.23\textwidth]{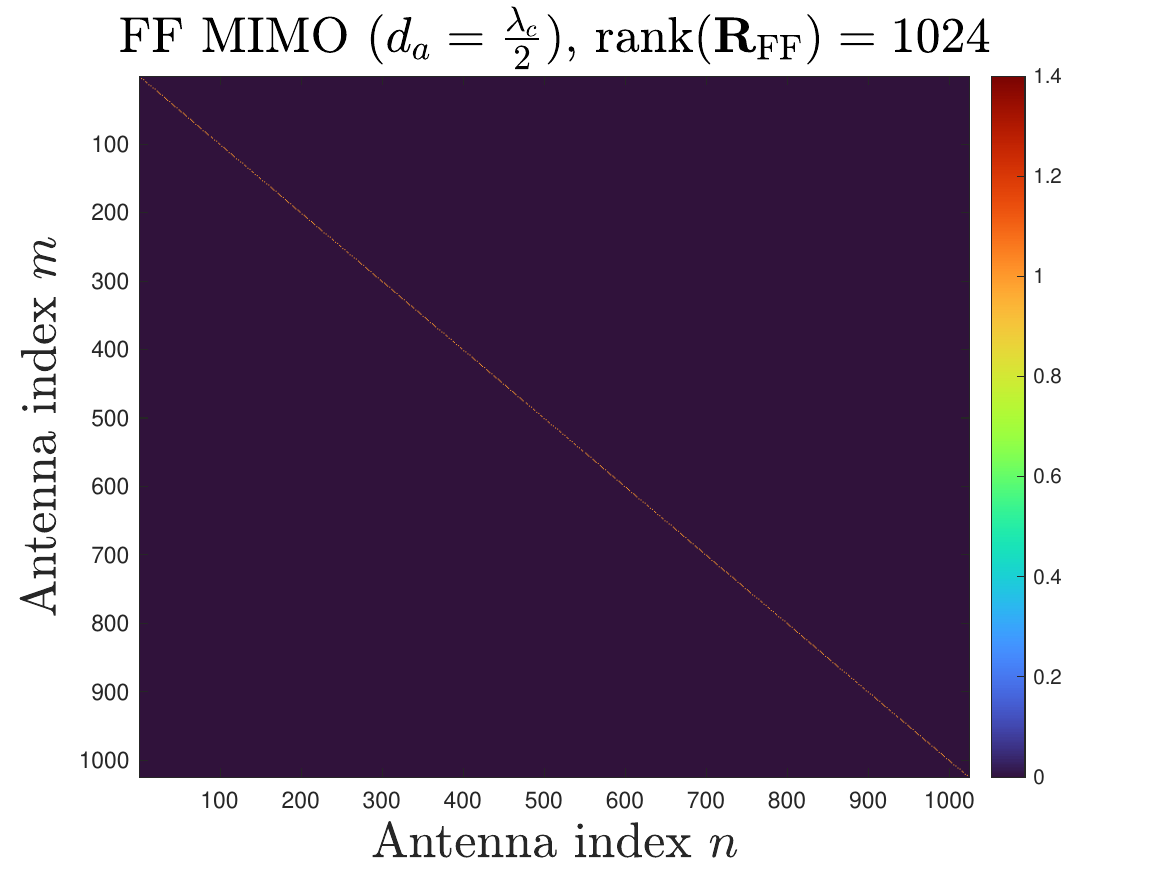}}\caption{\textcolor{black}{Comparison of the heat map of different spatial
correlation matrices. The BS is equipped with a ULA with 1024 antennas
that operates at a carrier frequency of $f_{c}=3.5$ GHz, with $S=20$
m, $R=3$ m, $\Psi=\frac{\pi}{3}$, $\mu=\frac{\pi}{4}$, and $\kappa=0$.
The following four cases are plotted: (a) Near-field HMIMO with $d_{a}=\frac{\lambda_{c}}{10}$,
(b) Far-field HMIMO with $d_{a}=\frac{\lambda_{c}}{10}$, (c) Near-field
MIMO with $d_{a}=\frac{\lambda_{c}}{2}$, (d) Far-field MIMO with
$d_{a}=\frac{\lambda_{c}}{2}$. The ranks of different spatial correlation
matrices are also labeled in the caption. \label{fig:Difference-FF-NF-correlation}}}
\end{figure}

By contrast, the far-field spatial correlation $\mathbf{R}_{\text{FF}}\in\mathbb{C}^{N\times N}$
is characterized by the following integration \cite{2023Damico,2022Dong},
i.e.,
\begin{equation}
\mathbf{R}_{\text{FF}}(n,m)=\beta_{0}\int_{-\pi}^{\pi}e^{-j\frac{2\pi}{\lambda_{c}}(m-n)d_{a}\sin\theta}f(\vartheta)\mathrm{d}\vartheta,\label{eq:FF-correlation-integral}
\end{equation}
in which $\vartheta$ is the angle of arrival (AoA), and $f(\vartheta)$
is the \textit{power angular spectrum (PAS)} characterizing the power
distribution over different AoAs \cite{2022Dong}. In particular,
in isotropic scattering environments where the multi-paths are uniformly
distributed, $\mathbf{R}_{\text{FF}}$ has a closed-form expression,
i.e., \cite{2022Andrea}
\begin{equation}
\mathbf{R}_{\text{FF-iso}}(n,m)=\operatorname{sinc}\left(\frac{2\left\Vert \mathbf{w}_{n}-\mathbf{w}_{m}\right\Vert }{\lambda_{c}}\right).
\end{equation}
Otherwise, in non-isotropic scattering environments, $\mathbf{R}_{\text{FF}}$
can also be obtained through numerical integration. 

The differences between the near-field and the far-field spatial correlations
are two-fold. First, $\mathbf{R}_{\text{FF}}$ is solely determined
by the PAS, $f(\vartheta)$, whereas $\mathbf{R}_{\text{NF}}$ relies
not only on the AoAs but also on the scatterers' positions, denoted
as the PLS, $f(\mathbf{s})$. Second, although $\mathbf{R}_{\text{FF}}$
exhibits spatial wide-sense stationarity, as evidenced by $\mathbf{R}_{\text{FF}}(n,m)$
being solely dependent on array index difference, $m-n$, this stationary
characteristic does not generally apply to $\mathbf{R}_{\text{NF}}$.
When $r(\mathbf{s})\gg Nd_{a},\forall\mathbf{s}\in\mathbf{S}$ holds
true, the two expressions are approximately equal, i.e., $\mathbf{R}_{\text{NF}}\approx\mathbf{R}_{\text{FF}}$,
but in other cases, the exact near-field correlation $\mathbf{R}_{\text{NF}}$
must be taken into account. For better illustration, we plot a visualized
comparison of $\mathbf{R}_{\text{NF}}$ and $\mathbf{R}_{\text{FF}}$
in Fig. \ref{fig:Difference-FF-NF-correlation}. It is observed that
the element of the far-field spatial correlation $\mathbf{R}_{\text{FF}}(n,m)$
is solely determined by the index difference $n-m$, which is called
the spatial stationarity property. By contrast, such a property does
not hold for the near-field spatial correlation, whose element $\mathbf{R}_{\text{NF}}(n,m)$
depends explicitly on both indexes $n$ and $m$. Also, the near-field
correlation $\mathbf{R}_{\text{NF}}$ results in a non-uniform average
power distribution among the ante\textcolor{black}{nnas. We also compare
the spatial correlation matrices for both the HMIMO ($d_{a}=\frac{\lambda_{c}}{10}$)
and the nominal MIMO systems ($d_{a}=\frac{\lambda_{c}}{2}$). It
is demonstrated that, in both the far-field and the near-field cases,
the HMIMO systems exhibit a much stronger spatial correlation compared
with the MIMO counterpart. The rank of the spatial correlation matrix
is the lowest in the near-field HMIMO case, indicating the strongest
correlation. }

\subsection{System Model for Near-Field HMIMO}

We introduce the system model for the near-field HMIMO. We mainly
consider two mainstream transceiver architectures, i.e., the fully-digital
and the hybrid analog-digital ones \cite{2019Zhang}. We assume orthogonal
pilots are utilized and hence focus on a specific user without loss
of generality. In the uplink channel estimation, the user sends pilots
to the BS for $K$ time slots. The received pilot signals, $\mathbf{\bar{y}}_{k}\in\mathbb{C}^{N_{\text{RF}}\times1}$,
at time slot $k$ is
\begin{equation}
\mathbf{\bar{y}}_{k}=\mathbf{W}_{k}^{H}(\bar{\mathbf{h}}s_{k}+\mathbf{\bar{n}}_{k}),
\end{equation}
where $\mathbf{W}_{k}\in\mathbb{C}^{N\times N_{\text{RF}}}$ is the
pilot combining matrix, $s_{k}=1$ is the pilot symbol, and $\mathbf{\bar{n}}_{k}\sim\mathcal{CN}(\mathbf{0},\sigma_{\mathbf{\bar{n}}}^{2}\mathbf{I})$
refers to the additive white Gaussian noise. After $K$ time slots
of pilot transmission, the received pilot signals are $\bar{\mathbf{y}}=\mathbf{\bar{M}}\mathbf{\bar{h}}+\mathbf{\bar{n}}$,
in which $\bar{\text{\ensuremath{\mathbf{M}}}}=[\mathbf{W}_{1}^{T},\mathbf{W}_{2}^{T},\ldots,\mathbf{W}_{K}^{T}]^{T}\in\mathbb{C}^{KN_{\text{RF}}\times N}$,
$\bar{\mathbf{y}}=[\mathbf{\bar{y}}_{1}^{T},\mathbf{\bar{y}}_{2}^{T},\ldots,\mathbf{\bar{y}}_{K}^{T}]^{T}\in\mathbb{C}^{KN_{\text{RF}}\times1}$,
and the noise $\mathbf{\bar{n}}$ is obtained similarly and whitened
as \cite{2023Yu}. In fully-digital HMIMO systems, the number of radio
frequency chains is equal to that of antennas, i.e., $N_{\text{RF}}=N$.
We may set the combiner as $\mathbf{W}_{k}=\mathbf{I}$, and the system
model reduces to $\bar{\mathbf{y}}=\mathbf{\bar{h}}+\mathbf{\bar{n}}$
\cite{2022Demir,2023An,2023Damico}. In contrast, in hybrid analog-digital
systems, the pilot combing matrix is the product of an \textcolor{black}{analog/holographic
combiner}\footnote{\textcolor{black}{The overall architecture of the hybrid analog-digital
transceivers are similar for the HMIMO \cite{2022Deng} and the MIMO
systems \cite{2016Yu}. The difference mainly lies in how the analog
combiners are implemented physically. Specifically, in the HMIMO systems,
the analog/holographic combiners are often implemented by low-cost
metasurfaces \cite{2022Deng}, while in the MIMO systems, the analog
combiners are mostly implemented by traditional phase shifters \cite{2016Yu}. }} $\mathbf{W}_{\text{RF},k}^{H}$ and a digital combiner $\mathbf{W}_{\text{BB},k}^{H}$,
i.e., $\mathbf{W}_{k}^{H}=\mathbf{W}_{\text{RF},k}^{H}\mathbf{W}_{\text{BB},k}^{H}$
\cite{2022Deng,2016Yu}. Notice that the combiners cannot be optimally
tuned without knowledge of the channel. Without loss of generality,
we consider an arbitrary scenario in which the digital combiner $\mathbf{W}_{\text{BB},k}^{H}$
is configured as identity, and the elements of the \textcolor{black}{analog/holographic
co}mbiner $\mathbf{W}_{\text{RF},k}^{H}$ are randomly selected from
one-bit \textcolor{black}{quantized angles} for the energy efficiency
considerations \cite{2023Yu-JSTSP,2020He}. The elements of $\bar{\text{\ensuremath{\mathbf{M}}}}$
\textcolor{black}{follow} the same distribution as $\mathbf{W}_{\text{RF},k}^{H}$,
that is $(\bar{\text{\ensuremath{\mathbf{M}}}})_{i,j}\in\frac{1}{\sqrt{N/N_{\text{RF}}}}\{\pm1\}$.
As most of the existing deep learning libraries handle real-valued
inputs, we transform the received pilot signals into a equivalent
real-valued form, 
\begin{equation}
\mathbf{y=\mathbf{Mh}+\mathbf{n}},\label{eq:hybrid-analog-digital-system-model}
\end{equation}
in which\textcolor{black}{{} $\mathbf{M}=\text{blkdiag}(\Re(\mathbf{\bar{M}}),\Im(\mathbf{\bar{M}}))\in\mathbb{R}^{2KN_{\text{RF}}\times2N}$},
$\mathbf{h}=[\Re(\mathbf{\bar{h}})^{T},\Im(\mathbf{\bar{h}})^{T}]^{T}\in\mathbb{R}^{2N}$,
$\mathbf{y}=[\Re(\mathbf{\bar{y}})^{T},\Im(\mathbf{\bar{y}})^{T}]^{T}\in\mathbb{R}^{2KN_{\text{RF}}}$,
$\mathbf{n}=[\Re(\mathbf{\bar{n}})^{T},\Im(\mathbf{\bar{n}})^{T}]^{T}\sim\mathcal{N}(\mathbf{0},\sigma_{\mathbf{\mathbf{n}}}^{2}\mathbf{I})$
with $\sigma_{\mathbf{\mathbf{n}}}^{2}=\frac{1}{2}\sigma_{\mathbf{\bar{n}}}^{2}$. 

Given the fact that $N_{\text{RF}}\ll N$ holds in hybrid analog-digital
systems and a small pilot overhead $K$, the measurement matrix $\mathbf{M}$
is often a fat matrix with $KN_{\text{RF}}<N$. This means that the
dimension of $\mathbf{y}$ is smaller than that of the channels $\mathbf{h}$,
which renders channel estimation an under-determined problem with
infinitely many possible solutions. This necessitates the use of the
prior knowledge of the channel $\mathbf{h}$ in some forms to reach
a high-quality solution. By contrast, since $N_{\text{RF}}=N$ in
fully-digital transceivers, it is good enough to simply set $K=1$
in the channel estimation stage. In such a scenario, $\mathbf{M}$
is a square matrix, i.e., $\mathbf{M=I}$, and the system model becomes
\begin{equation}
\mathbf{y=\mathbf{h}+\mathbf{n}},\label{eq:fully-digital-system-model}
\end{equation}
which reduces to a denoising problem. Nevertheless, to achieve competitive
estimation performance, it is essential to leverage the prior of the
channel $\mathbf{h}$, not to mention the Bayes-optimal MMSE estimator,
which typically requires full knowledge of the prior distribution
$p(\mathbf{h})$. 

However, the grand challenge is that, we actually have very limited
prior knowledge of the channel $\mathbf{h}$ in real environments,
let alone characterizing its full prior distribution. As mentioned
in Section \ref{sec:Introduction}, previous works either attempt
to adopt simplified priors, e.g., based on sparsity and low-rankness,
or introduce the postulated ones, e.g., by using mixture distributions,
which, however, can only characterize limited aspects of the channel
feature, and hence often result in sub-optimal performance. On the
other hand, existing DL methods mostly rely on a supervised dataset
of ground-truth channels $\mathbf{h}$, which are hardly available
in practice. Hence, in the next section, we study how to achieve \textcolor{black}{Bayes-optimal}
channel estimation without priors or supervision, which is extremely
important in practice. 

\section{General Frameworks of the Score-Based Channel Estimator\label{sec:Bayes-Optimal-Unsupervised-Chann}}

In this section, we propose general frameworks and enabling techniques
of the Bayes-optimal score-based channel estimator. We first consider
the fully-digital transceivers and derive the closed-form expression
of the Bayes-optimal channel estimator solely based on the received
pilot signals $\mathbf{y}$. Next, we propose techniques to acquire
the two crucial elements of the estimator, namely the score function
and the noise level, based on score matching and PCA, respectively. 

\subsection{Bridging MMSE Estimation with the Score Function\label{subsec:Bridging-MMSE-Estimation}}

In this subsection, we first focus on the fully-digital system model,
i.e., (\ref{eq:fully-digital-system-model}). Extension to hybrid
analog-digital HMIMO systems will be presented in Section \ref{subsec:Extension-to-Hybrid}. 

Our target is a Bayes-optimal channel estimator, $\mathbf{\hat{h}}=D(\mathbf{y})$,
that minimizes the mean-square-error (MSE), i.e.,
\begin{equation}
\text{MSE}\triangleq\text{\ensuremath{\mathbb{E}(\|\mathbf{h}-\hat{\mathbf{h}}\|^{2}|\mathbf{y})=\int\|\mathbf{h}-\mathbf{\hat{h}}\|^{2}p(\mathbf{h}|\mathbf{y})d\mathbf{h}}},
\end{equation}
where the expectation above is taken with respect to (w.r.t.) the
unknown channel $\mathbf{h}$, while $p(\mathbf{h}|\mathbf{y})$ is
the posterior density. Taking the derivative of the above equation
w.r.t. $\hat{\mathbf{h}}$ and nulling it, we reach the Bayes-optimal,
i.e., minimum MSE (MMSE), channel estimator, given by
\begin{equation}
\begin{aligned}\hat{\mathbf{h}}_{\text{MMSE}} & =\mathbb{E}(\mathbf{h}|\mathbf{y})=\int\mathbf{h}p(\mathbf{h}|\mathbf{y})d\mathbf{h}=\int\mathbf{h}\frac{p(\mathbf{h},\mathbf{y})}{p(\mathbf{y})}d\mathbf{h},\end{aligned}
\label{eq:posterior-mean}
\end{equation}
where $p(\mathbf{h},\mathbf{y})$ is the joint density, and $p(\mathbf{y})$
is the measurement density obtained via marginalization, i.e., 
\begin{equation}
\begin{aligned}p(\mathbf{y}) & =\int p(\mathbf{h},\mathbf{y})d\mathbf{h}=\int p(\mathbf{y}|\mathbf{h})p(\mathbf{h})d\mathbf{h}\\
 & =\left(\frac{1}{2\pi\sigma_{\mathbf{n}}^{2}}\right)^{N}\int e^{-\frac{1}{2\sigma_{\mathbf{n}}^{2}}\|\mathbf{y}-\mathbf{h}\|^{2}}p(\mathbf{h})d\mathbf{h}.
\end{aligned}
\label{eq:measurement-density}
\end{equation}

The last equality holds since the likelihood function $p(\mathbf{y}|\mathbf{h})\sim\mathcal{N}(\mathbf{h},\sigma_{\mathbf{n}}^{2}\mathbf{I})$
expresses $p(\mathbf{y})$ as a convolution between the prior distribution
$p(\mathbf{h})$ and the i.i.d. Gaussian noise. Taking the derivative
of both sides of (\ref{eq:measurement-density}) w.r.t. $\mathbf{y}$
gives 
\begin{equation}
\begin{aligned}\nabla_{\mathbf{y}}p(\mathbf{y}) & =\left(\frac{1}{2\pi\sigma_{\mathbf{n}}^{2}}\right)^{N}\int\nabla_{\mathbf{y}}e^{-\frac{1}{2\sigma_{\mathbf{n}}^{2}}\|\mathbf{y}-\mathbf{h}\|^{2}}p(\mathbf{h})d\mathbf{h}\\
 & =\frac{1}{\sigma_{\mathbf{n}}^{2}}\left(\frac{1}{2\pi\sigma_{\mathbf{n}}^{2}}\right)^{N}\int(\mathbf{h}-\mathbf{y})e^{-\frac{1}{2\sigma_{\mathbf{n}}^{2}}\|\mathbf{y}-\mathbf{h}\|^{2}}p(\mathbf{h})d\mathbf{h}\\
 & =\frac{1}{\sigma_{\mathbf{n}}^{2}}\int(\mathbf{h}-\mathbf{y})p(\mathbf{y}|\mathbf{h})p(\mathbf{h})d\mathbf{h}.
\end{aligned}
\label{eq:gradient-py}
\end{equation}
Dividing both sides of (\ref{eq:gradient-py}) w.r.t. $p(\mathbf{y})$
results in the following: 
\begin{equation}
\begin{aligned}\frac{\nabla_{\mathbf{y}}p(\mathbf{y})}{p(\mathbf{y})} & =\frac{1}{\sigma_{\mathbf{n}}^{2}}\int(\mathbf{h}-\mathbf{y})\frac{p(\mathbf{y}|\mathbf{h})p(\mathbf{h})}{p(\mathbf{y})}d\mathbf{h}\\
 & =\frac{1}{\sigma_{\mathbf{n}}^{2}}\int(\mathbf{h}-\mathbf{y})p(\mathbf{h}|\mathbf{y})d\mathbf{h}\\
 & =\frac{1}{\sigma_{\mathbf{n}}^{2}}\int\mathbf{h}p(\mathbf{h}|\mathbf{y})d\mathbf{h}-\frac{1}{\sigma_{\mathbf{n}}^{2}}\mathbf{y}\int p(\mathbf{h}|\mathbf{y})d\mathbf{h}\\
 & =\frac{1}{\sigma_{\mathbf{n}}^{2}}\left(\mathbf{\hat{h}}_{\text{MMSE}}-\mathbf{y}\right)\text{,}
\end{aligned}
\end{equation}
where the second equality holds owing to the Bayes\textquoteright{}
theorem, and the last equality holds due to (\ref{eq:posterior-mean})
and $\int p(\mathbf{h}|\mathbf{y})d\mathbf{h}=1$. By rearranging
the terms and plugging in $\frac{\nabla_{\mathbf{y}}p(\mathbf{y})}{p(\mathbf{y})}=\nabla_{\mathbf{y}}\log p(\mathbf{y})$,
we reach the foundation of the proposed algorithm: 
\begin{equation}
\hat{\mathbf{h}}_{\text{MMSE}}=\mathbf{y}+\sigma_{\mathbf{n}}^{2}\nabla_{\mathbf{y}}\log p(\mathbf{y}),\label{eq:Tweedie}
\end{equation}
in which $\nabla_{\mathbf{y}}\log p(\mathbf{y})$ is called the \textit{Stein's
score function} in statistics \cite{2014Alain}. Based on (\ref{eq:Tweedie}),
we notice that the Bayes-optimal MMSE channel estimator could be achieved
\textit{solely based on the received pilot signals} $\mathbf{y}$,
without having access to the prior distribution $p(\mathbf{h})$ or
a supervised training dataset consisting of the ground-truth channels
$\mathbf{h}$, which are both unavailable in practice. With efficient
estimators of the noise level $\sigma_{\mathbf{n}}$ and the score
function $\nabla_{\mathbf{y}}\log p(\mathbf{y})$, the Bayes-optimal
MMSE channel estimator can be computed in a closed form with extremely
low complexity. In the following, we discuss how to utilize the machine
learning tools to obtain an accurate estimation of them based\textit{
solely} on $\mathbf{y}$. 
\begin{rem}
The above derivation depends on the nice properties of the additive
white Gaussian noise $\mathbf{n}$, i.e., $\mathbf{n}\sim\mathcal{CN}(\mathbf{0},\sigma_{\mathbf{n}}^{2}\mathbf{I})$,
which is the most commonly seen noise distribution in MIMO systems
and has been widely adopted in the previous works on HMIMO channel
estimation \cite{2022Demir,2023An,2023Damico}. Nevertheless, one
can easily see that it also holds when the noise follows correlated
Gaussian distribution\footnote{The correlated noise could be observed as electromagnetic interference
in holographic RIS-assisted wireless systems \cite{2022Andrea}. }, e.g., $\mathbf{n}\sim\mathcal{CN}(\mathbf{0},\bm{\Sigma})$, by
replacing $\sigma_{\mathbf{n}}^{2}$ in (\ref{eq:Tweedie}) with the
covariance matrix $\bm{\Sigma}$. Furthermore, when the environment
contains some long-tailed Cauchy noise \cite{2023Gulgun}, the extension
can also be derived using a similar procedure \cite{2011Raphan}.
We omit further discussion here due to the limited space. 
\end{rem}

\subsection{Unsupervised Learning of the Score Function}

We discuss how to get the score function $\nabla_{\mathbf{y}}\log p(\mathbf{y})$.
Given that a closed-form expression is intractable to acquire, we
instead aim to achieve a parameterized function with a neural network,
and discuss how to train it based on score matching. We first introduce
the denoising anto-encoder (DAE) \cite{2014Alain}, the core of the
training process, and explain how to utilize it to approximate the
score function. 

To obtain the score function $\nabla_{\mathbf{y}}\log p(\mathbf{y})$,
the measurement $\mathbf{y}$ is treated as the target signal that
the DAE should denoise. The general idea is to obtain the score function
$\nabla_{\mathbf{y}}\log p(\mathbf{y})$ based on its analytical relationship
with the DAE of $\mathbf{y}$, which will be established later in
\textbf{Theorem \ref{thm:The-optimal-DAE,}}. We first construct a
noisy version of the target signal $\mathbf{y}$ by \textit{manually}
adding some additive white Gaussian noise, $\text{\ensuremath{\varsigma}}\mathbf{u}$,
where $\mathbf{u}\sim\mathcal{N}(\mathbf{0},\mathbf{I})$ and $\varsigma$
controls the noise level\footnote{Note that the extra noise is only added during the training process. },
and then train a DAE to \textit{denoise} the manually added noise.
The DAE, denoted by $R_{\bm{\theta}}(\cdot;\cdot)$, is trained by
the $\ell_{2}$-loss function $\mathbb{E}\|\mathbf{y}-R_{\bm{\theta}}(\mathbf{y};\varsigma)\|^{2}$.
\textbf{Theorem 2} explains the relationship of the score function
and the DAE. 
\begin{thm}[{Alain-Bengio \cite[Theorem 1]{2014Alain}}]
\label{thm:The-optimal-DAE,}The optimal DAE, $R_{\bm{\theta}^{*}}(\cdot;\cdot)$,
behaves asymptotically as
\begin{equation}
R_{\bm{\theta}^{*}}(\mathbf{y};\varsigma)=\mathbf{y}+\varsigma^{2}\nabla_{\mathbf{y}}\log p(\mathbf{y})+o(\varsigma^{2}),\,\,\text{\text{as}}\,\,\varsigma\rightarrow0.\label{eq:optimal DAE}
\end{equation}
\end{thm}
\begin{IEEEproof}
Please refer to \cite[Appendix A]{2014Alain}. 
\end{IEEEproof}
The above theorem illustrates that, for a sufficiently small $\varsigma$,
we can approximate the score function based on the DAE by $\nabla_{\mathbf{y}}\log p(\mathbf{y})\thickapprox\frac{R_{\bm{\theta}}(\mathbf{y};\varsigma)-\mathbf{y}}{\varsigma^{2}}$,
assuming that parameter of the DAE, $\bm{\theta}$, is near-optimal,
i.e., $\bm{\theta}\thickapprox\bm{\theta}^{*}$. Nevertheless, the
approximation can be numerically unstable as the denominator, $\text{\ensuremath{\varsigma^{2}}}$,
is close to zero. To alleviate the problem, we improve the structure
of the DAE and rescale the original loss function. 

First, we consider a residual form of the DAE with a scaling factor.
Specifically, let $R_{\bm{\theta}}(\mathbf{y};\varsigma)=\varsigma^{2}S_{\bm{\theta}}(\mathbf{y};\varsigma)+\mathbf{y}$.
Plugging it into (\ref{eq:optimal DAE}), the score function is approximately
equal to 
\begin{equation}
\nabla_{\mathbf{y}}\log p(\mathbf{y})\thickapprox\frac{(\varsigma^{2}S_{\bm{\theta}}(\mathbf{y};\varsigma)+\mathbf{y})-\mathbf{y}}{\varsigma^{2}}=S_{\bm{\theta}}(\mathbf{y};\varsigma),\,\,\text{\text{as}}\,\,\varsigma\rightarrow0.\label{eq:approximate-AR-DAE}
\end{equation}
This reparameterization enables $S_{\bm{\theta}}(\mathbf{y};\varsigma)$
to approximate the score function directly, thereby circumventing
the requirement for division that may lead to numerical instability.
In addition, the residual link significantly enhances the denoising
capability of DAE, as it can easily learn an identity mapping \cite{2016He}. 

Second, since the variance $\varsigma^{2}$ of the manually added
noise is small, the gradient of the original $\ell_{2}$-loss function,
i.e., $\mathbb{E}\|\mathbf{y}-R_{\bm{\theta}}(\mathbf{y};\varsigma)\|^{2}$,
can easily vanish to zero and may further lead to difficulties in
training. Hence, we rescale the loss function by a factor of $\frac{1}{\varsigma}$
to safeguard the vanishing gradient issue, i.e., 
\begin{equation}
\text{\ensuremath{\mathcal{L}_{\text{DAE}}(\bm{\theta})=\mathbb{E}\|\mathbf{u}+\varsigma S_{\bm{\theta}}(\mathbf{y};\varsigma)\|^{2}}},\label{eq:improved-DAE-loss}
\end{equation}
where (\ref{eq:approximate-AR-DAE}) is plugged into the loss function.
\textcolor{black}{The expectation is taken with respect to $\mathbf{y}$
and $\text{\ensuremath{\mathbf{u}}}$. }

We are interested in the region where $\varsigma$ is sufficiently
close to zero, in which case $S_{\bm{\theta}}(\mathbf{y};0)$ can
be deemed equal to the score function $\nabla_{\mathbf{y}}\log p(\mathbf{y})$
according to (\ref{eq:approximate-AR-DAE}). Nevertheless, directly
training the network using a very small $\varsigma$ is difficult
since the SNR of the gradient signal decreases in a linear rate $\mathcal{O}(\varsigma)$
with respect to $\varsigma$, which introduces difficulty for the
stochastic gradient descent \cite{2020Lim}. To exploit the asymptotic
optimality of the score function approximation when $\varsigma\rightarrow0$,
we propose to simultaneously train a network $S_{\bm{\theta}}(\mathbf{y};\varsigma)$
conditioned on varying $\varsigma$ values in each epoch, such that
it can handle various $\varsigma$ levels and generalize to the desired
region $\varsigma\rightarrow0$, i.e., $S_{\bm{\theta}}(\mathbf{y};0)$.
To achieve the goal, we linearly anneal the level of the added noise
$\varsigma\in[\varsigma_{\text{min}},\varsigma_{\text{max}}]$ from
a large value to a small one in each epoch. That is, we \textit{condition}
$S_{\bm{\theta}}(\mathbf{y};\varsigma)$ on the manually added noise
level $\varsigma$ during training. 

The proposed algorithm is described in \textbf{Algorithm 1}. The DAE
is trained with a stochastic gradient descent for $Q$ epochs. In
each epoch, we draw a random vector $\mathbf{u}$ and anneal $\varsigma$
to control the extra noise level according to the current number of
iterations $q$. Then, the scaled DAE loss function $\mathcal{L}_{\text{DAE}}$
in (\ref{eq:improved-DAE-loss}) is minimized by stochastic gradient
descent. Note that in the training process, nothing but a dataset
of the received pilot signals $\mathbf{y}$ is necessary, which is
readily available in practice. In the inference stage, one can apply
formula (\ref{eq:Tweedie}) to compute the score-based MMSE estimator,
in which the score function can be approximated by using $S_{\bm{\theta}}(\mathbf{y};0)$,
i.e., setting $\varsigma$ as zero, and noise level $\sigma_{\mathbf{\bar{n}}}$
($\sigma_{\mathbf{\bar{n}}}=\sqrt{2}\sigma_{\mathbf{n}}$) can be
estimated by the PCA-based algorithm in the next subsection. 

\floatplacement{algorithm}{t}
\begin{algorithm} 
\caption{Training the score function estimator}
\begin{algorithmic}[1]

\STATE {\bf Input:} Learning rate $\gamma$, maximum extra noise level $\varsigma_{\text{max}}$, minimum extra noise level $\varsigma_{\text{min}}$, number of epochs $Q$, a dataset of received pilot signals $\{\mathbf{y}_i\}_{i=1}^M$ \\
\STATE {\bf Output:} Trained DAE parameters $\bm{\theta}$ \\
\STATE {\bf for} $q=1:Q$ {\bf do}
\STATE\hspace{\algorithmicindent} Draw $\mathbf{u} \sim \mathcal{N}(\mathbf{0},\mathbf{I})$
\STATE\hspace{\algorithmicindent} Set the extra noise level with $\varsigma \gets \frac{Q-q}{Q}\varsigma_{\text{min}}+\frac{q}{Q}\varsigma_{\text{max}}$
\STATE\hspace{\algorithmicindent} Compute the loss function $\mathcal{L}_\text{DAE}$ as in (18)
\STATE\hspace{\algorithmicindent} Update DAE parameters by $\bm{\theta} \gets \bm{\theta}-\gamma \nabla_{\bm{\theta}}\mathcal{L}_\text{DAE}$
\STATE {\bf return} $\bm{\theta}$  \\
\end{algorithmic}
\end{algorithm}

\subsection{PCA-Based Noise Level Estimation}

We propose a low-complexity PCA-based algorithm to estimate the noise
level $\sigma_{\mathbf{n}}$ required in (\ref{eq:Tweedie}) based
on \textit{a single instance} of the received pilot signals $\mathbf{y}\sim\mathcal{N}(\mathbf{h},\sigma_{\mathbf{n}}^{2}\mathbf{I})$,
while the mean $\mathbf{h}$, i.e., the ground-truth HMIMO channel,
is completely unknown. Mathematically, the problem is to estimate
$\sigma_{\mathbf{n}}$ from a single realization of a multivariate
Gaussian distribution with unknown mean, which is certainly an intractable
problem if the mean, i.e., the HMIMO channel, does not have any particular
structure. In the following, we first motivate the idea by analyzing
the HMIMO spatial correlation matrix, and then present the proposed
PCA-based algorithm, i.e., \textbf{Algorithm 2}. 

The basic idea behind is the low-rank property of the spatial correlation
matrix $\mathbf{R}$ of HMIMO due to the dense deployment of the antenna
elements, which introduces a strong correlation between adjacent antenna
elements. For example, for far-field isotropic scattering environments,
the rank of the correlation matrix $\mathbf{R}_{\text{FF-iso}}$ is
roughly $\text{rank}(\mathbf{R}_{\text{FF-iso}})\thickapprox\pi Nd_{a}^{2}/\lambda_{c}^{2}$
\cite{2022Pizzo}. It decreases with the shrink of antenna spacing
and the increase of the carrier frequency. For example, when $d_{a}=\lambda_{c}/4$,
around 80\% of the eigenvalues of $\mathbf{R}_{\text{FF-iso}}$ shrink
towards zero. The rank deficiency phenomenon of the spatial correlation
matrix tends to be even more prominent in non-isotropic scattering
environments \cite{2023An}, particularly in the near-field region
\cite{2023Lu}. For example, for the same near-field system configurations
as Fig. \ref{fig:Difference-FF-NF-correlation}(a), more than 95\%
of the eigenvalues of $\mathbf{R}_{\text{NF}}$ are close to zero.
Hence, we propose to make use of the low-rankness of the HMIMO spatial
correlation to estimate $\sigma_{\mathbf{n}}$. \begin{algorithm}[!tbp]
\caption{PCA-based noise level estimation}
\begin{algorithmic}[1]
\STATE {\bf Input:} Received pilot signals ${\mathbf{y}} \in \mathbb{R}^{N \times 2}$, the size of the sliding window $d$\\
\STATE {\bf Output:} Estimated noise level $\hat{\sigma}_{\bar{\mathbf{n}}}$ $(\hat{\sigma}_{\bar{\mathbf{n}}}=\sqrt{2}\hat{\sigma}_{{\mathbf{n}}})$\\
\STATE \textcolor{black}{Reshape $\mathbf{y}$ into the tensor form $\mathbf{Y} \in \mathbb{R}^{\sqrt{N} \times \sqrt{N} \times 2}$} \\
\STATE \textcolor{black}{Decompose $\mathbf{Y}$ into virtual subarrays $\{\mathbf{y}_t \in \mathbb{R}^{2d \times 1}\}_{t=1}^{s}$ by the sliding window, in which $s=(\sqrt{N}-\sqrt{d}+1)^2$} \\
\STATE \textcolor{black}{Calculate the mean vector of $\{\mathbf{y}_t \in \mathbb{R}^{2d \times 1}\}_{t=1}^{s}$, i.e., $\bm{\mu}=\frac{1}{s} \sum_{t=1}^{s}\mathbf{y}_{t} \in \mathbb{R}^{2d \times 1}$} \\
\STATE \textcolor{black}{Calculate the covariance matrix of $\{\mathbf{y}_t \in \mathbb{R}^{2d \times 1}\}_{t=1}^{s}$, i.e., $\bm{\Sigma}=\frac{1}{s}\sum_{t=1}^{s}(\mathbf{y}_{t}-\bm{\mu})(\mathbf{y}_{t}-\bm{\mu})^T \in \mathbb{R}^{2d \times 2d}$} \\
\STATE \textcolor{black}{Calculate the eigenvalues $\{\lambda_i\}_{i=1}^{2d}$ of the covariance $\bm{\Sigma}$ in descending order, i.e., $\lambda_1 \geq \lambda_2 \geq \ldots \geq \lambda_{2d}$} \\
\STATE \textcolor{black}{Iteratively split the eigenvalues $\{\lambda_i\}_{i=1}^{2d}$ into the principal and the redundant dimensions, and estimate the noise level based on the latter: } \\
\STATE {\bf for} $i = 1:2d$ {\bf do}
\STATE\hspace{\algorithmicindent} $\tau = \frac{1}{2d-i+1}\sum_{j=i}^{2d} \lambda_j$
\STATE\hspace{\algorithmicindent} {\bf if} $\tau$ is the median of $\{\lambda_j\}_{j=i}^{2d}$ {\bf then}
\STATE\hspace{\algorithmicindent}\hspace{\algorithmicindent}\hspace{\algorithmicindent} $\hat{\sigma}_{\bar{\mathbf{n}}}=\sqrt{2\tau}$ and {\bf break} 
\STATE\hspace{\algorithmicindent} {\bf end if }
\STATE {\bf end for}
\end{algorithmic}
\end{algorithm}

\begin{figure}[t]
\centering{}\subfloat[]{\centering{}\includegraphics[width=0.25\textwidth]{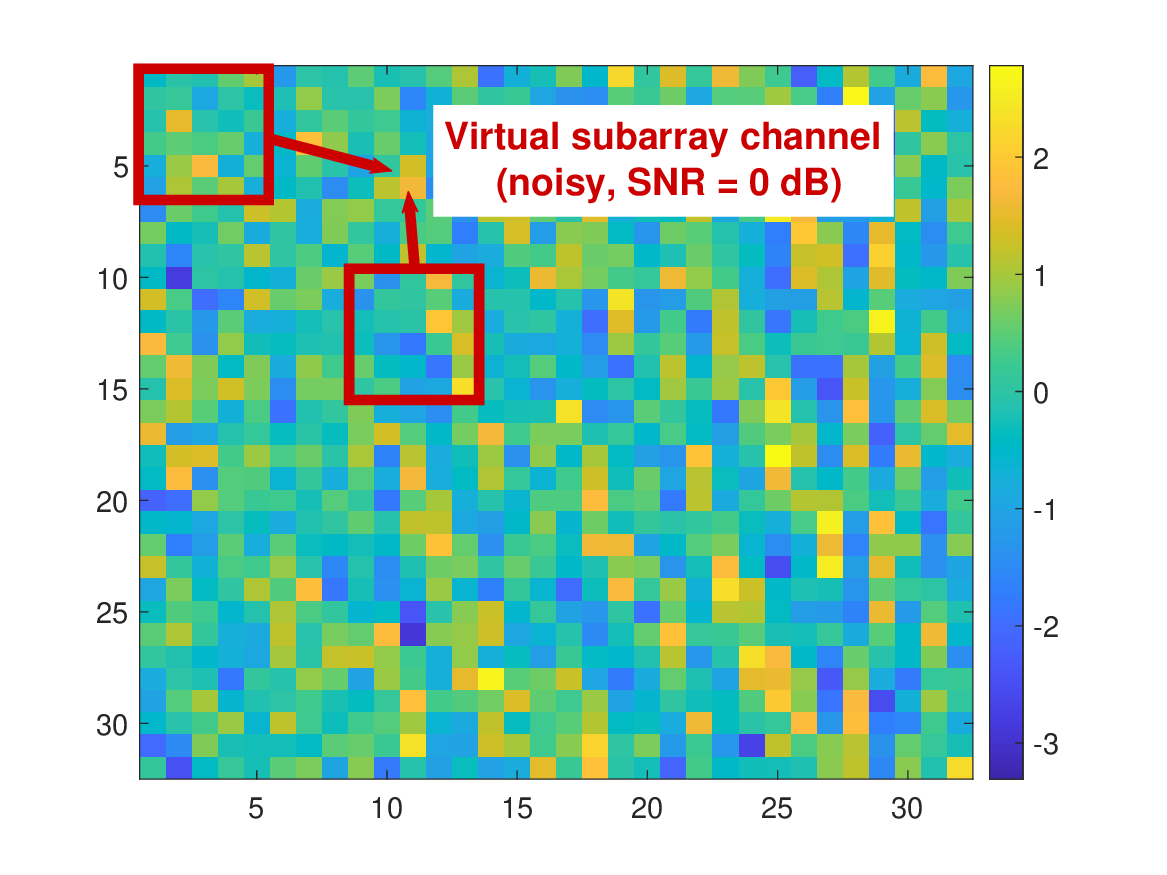}}\subfloat[]{\centering{}\includegraphics[width=0.25\textwidth]{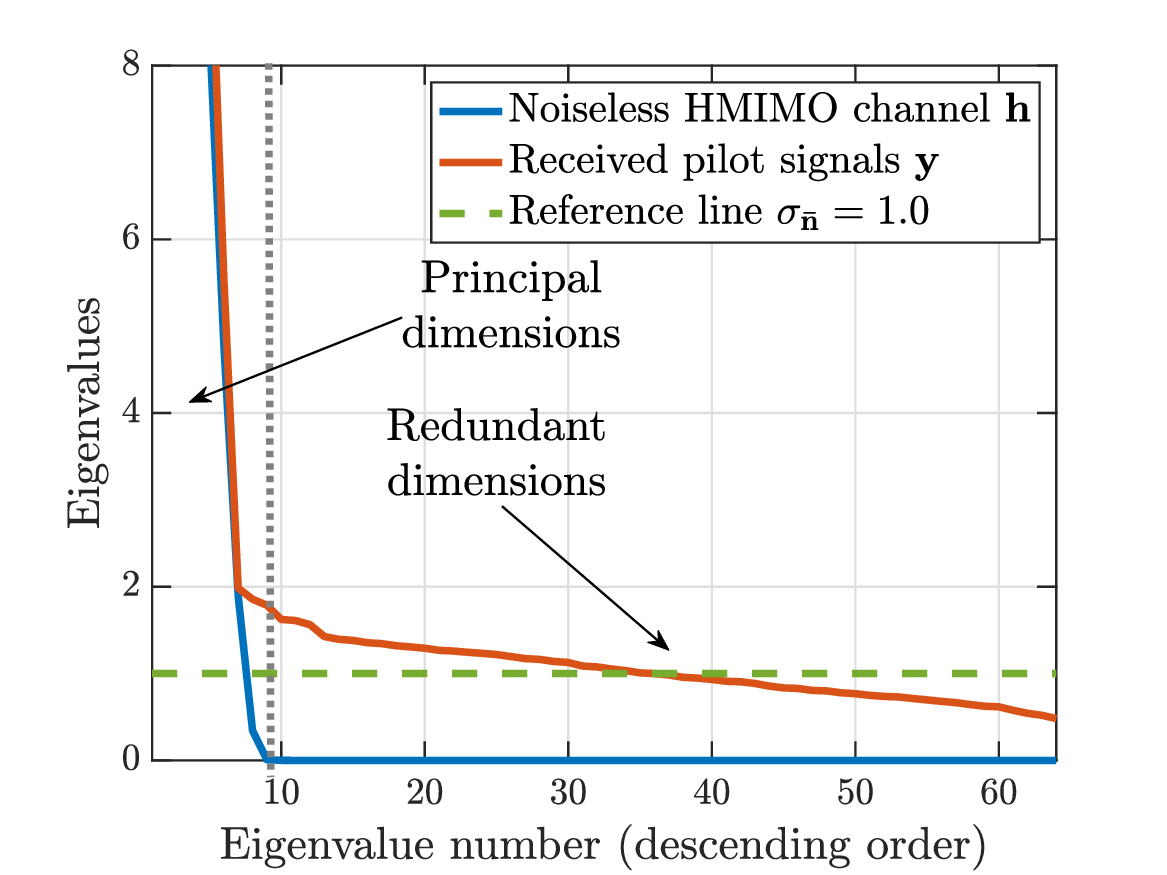}}\caption{(a) The heat map of the received pilot signals of the near-field HMIMO
channel when the received SNR is 0 dB. The other parameters are set
the same as Fig. \ref{fig:Difference-FF-NF-correlation}(a). (b) Eigenvalues
of the covariance matrices of the virtual subarray channels $\{\mathbf{h}_{t}\in\mathbb{R}^{2d\times1}\}_{t=1}^{s}$
and the pilots $\{\mathbf{y}_{t}\in\mathbb{R}^{2d\times1}\}_{t=1}^{s}$,
in which the principal and redundant dimensions are separated by a
dotted line. \label{fig:Eigenvalues-of-the}}
\end{figure}

We present the PCA-based algorithm by using an illustrative example.
Following the same setting as Fig. \ref{fig:Difference-FF-NF-correlation}(a),
we generate an HMIMO channel sample $\mathbf{h}$ and the corresponding
received pilot signals $\mathbf{y}=\mathbf{h}+\mathbf{n}$ when the
received SNR is 0 dB. We reshape $\mathbf{y},\mathbf{h},\mathbf{n}\in\mathbb{R}^{2N\times1}$
into their corresponding square tensors $\mathbf{Y},\mathbf{H},\mathbf{N}\in\mathbb{R}^{\sqrt{N}\times\sqrt{N}\times2}$,
and plotted the heat map of $\mathbf{Y}(:,:,1)$ as shown in Fig.
\ref{fig:Eigenvalues-of-the}(a). We then decompose $\mathbf{Y}$
into $s=(\sqrt{N}-\sqrt{d}+1)^{2}$ virtual subarray tensors by using
a sliding window of shape $\sqrt{d}\times\sqrt{d}\times2$, and then
reshape them back into $s$ vectors, i.e., $\{\mathbf{y}_{t}\in\mathbb{R}^{2d\times1}\}_{t=1}^{s}$.
Similarly, the noiseless channel $\mathbf{H}$ and the noise $\mathbf{N}$
can also be decomposed as $\{\mathbf{y}_{t}\in\mathbb{R}^{2d\times1}\}_{t=1}^{s}$
and $\{\mathbf{n}_{t}\in\mathbb{R}^{2d\times1}\}_{t=1}^{s}$, and
should satisfy 
\begin{equation}
\mathbf{y}_{t}=\mathbf{h}_{t}+\mathbf{n}_{t}.\label{eq:virtual-subarray-samples}
\end{equation}

Thanks to the low-rank property of the near-field spatial correlation
$\mathbf{R}_{\text{NF}}$, the decomposed virtual subarray channels
$\{\mathbf{h}_{t}\in\mathbb{R}^{2d\times1}\}_{t=1}^{s}$ should also
lie in a low-dimensional subspace. In Fig. \ref{fig:Eigenvalues-of-the}(b),
we plot the eigenvalues of the covariance matrices of the received
pilot signals $\{\mathbf{y}_{t}\in\mathbb{R}^{2d\times1}\}_{t=1}^{s}$
corresponding to the decomposed virtual subarray channels $\{\mathbf{h}_{t}\in\mathbb{R}^{2d\times1}\}_{t=1}^{s}$
in descending order, with a reference line that marks the noise level
$\sigma_{\mathbf{\bar{n}}}=\sqrt{2}\sigma_{\mathbf{n}}$. As observed
in the figure, the eigenvalues of $\{\mathbf{h}_{t}\in\mathbb{R}^{2d\times1}\}_{t=1}^{s}$
shrink to zero with only about 10 principal dimensions. The zero eigenvalues
correspond to the redundant dimensions. By contrast, we observe that
the eigenvalues of the covariance of $\{\mathbf{y}_{t}\in\mathbb{R}^{2d\times1}\}_{t=1}^{s}$
are concentrated around $\sigma_{\mathbf{\bar{n}}}$ in the redundant
dimensions. This example suggests that it is possible to estimate
the noise level based on the redundant eigenvalues, which can be shown
to follow a Gaussian distribution by using the following theorem.
The above manipulations correspond to lines 3 to 7 in \textbf{Algorithm
2}. 
\begin{thm}
The redundant eigenvalues of the covariance of $\{\mathbf{y}_{t}\in\mathbb{R}^{2d\times1}\}_{t=1}^{s}$
is Gaussian distributed with variance $\sigma_{\mathbf{\mathbf{n}}}^{2}$.
\label{thm:The-redundant-eigenvalues}
\end{thm}
\begin{IEEEproof}
Please refer to Appendix \ref{sec:Proof-of-Theorem-Gaussian}. 
\end{IEEEproof}
The above theorem motivates us to first separate the principal and
the redundant dimensions, and then estimate the noise level $\sigma_{\mathbf{\bar{n}}}$
based on the redundant ones. In fact, the principal and redundant
eigenvalues could be separated by an iterative process. Assuming that
the eigenvalue set $\text{\ensuremath{\mathbf{S}}}=\{\lambda_{i}\}_{i=1}^{2d}$
could be divided into two subsets $\mathbf{S}_{p}$ and $\mathbf{S}_{r}$
such that $\mathbf{S}=\mathbf{S}_{p}\text{\ensuremath{\cup\mathbf{S}_{r}}}$,
where $\mathbf{S}_{p}$ is the set of the principal eigenvalues and
$\mathbf{S}_{r}$ consists of the redundant eigenvalues. We first
assume that $\mathbf{S}_{p}=\varnothing$ and $\mathbf{S}_{r}=\mathbf{S}$,
and iteratively move elements from $\mathbf{S}_{r}$ to $\mathbf{S}_{p}$
as long as the condition that, the mean of $\mathbf{S}_{r}$ is also
the median of it, is not satisfied, as realized by lines 8 to 14 in
\textbf{Algorithm 2}\footnote{If the 'if' statement in line 11 is never satisfied, the estimated
noise is set as $\hat{\sigma}_{\bar{n}}=\sqrt{2\tau}$ when $i=2d$,
corresponding to the smallest eigenvalue $\lambda_{2d}$. }. If the redundant eigenvalues, following a Gaussian distribution,
could be successfully sorted out, we can then simply resort to the
sample standard deviation, which is the minimum variance unbiased
estimator, to estimate the noise level $\sigma_{\mathbf{\bar{n}}}$.
It has been proved in \cite{1965Shapiro} that, such a method offers
an accurate estimate of the noise level, as long as the principal
dimension is not too large compared to the overall dimension, which
can be easily satisfied by the near-field HMIMO channel since the
rank of the spatial correlation matrices is extremely low. Empirically,
we conclude that adopting a $5\times5$ sliding window is enough to
offer an accurate estimate when $N=1024$. More experimental evaluations
will be presented in Section \ref{sec:Simulation-Results}. 

\section{Practical Algorithms of the Score-Based Channel Estimator\label{sec:Practical-Algorithms-of}}

In this section, we present and analyze practical algorithms for the
proposed score-based channel estimator. We first focus on the fully-digital
systems in Section \ref{subsec:Score-Based-Estimator-in-fully-digital},
and then discuss how to extend it to hybrid analog-digital transceivers
based on iterative message passing algorithms in Section \ref{subsec:Extension-to-Hybrid}.
Lastly, we analyze the computational complexity. 

\subsection{Score-Based Estimator in Fully-Digital Systems\label{subsec:Score-Based-Estimator-in-fully-digital}}

The procedure of the score-based estimator in fully-digital near-field
HMIMO systems is summarized in \textbf{Algorithm 3}. In the offline
stage, the BS first collects a dataset of received pilot signals $\mathbf{y}$
to train the DAE neural network $\bm{\theta}$ by \textbf{Algorithm
1}. This will cost no additional overhead for the spectral resources,
and also does not require knowledge of the noise level. At the online
inference stage, \textbf{Algorithm 3} requires the received pilot
signals $\mathbf{y}$, the well-trained DAE parameters $\bm{\theta}$,
and the shape of the sliding window for estimating the noise level
$\sigma_{\mathbf{\bar{n}}}$. As an initialization, we first employ
the trained DAE network $\bm{\theta}$ to approximate the score function
using $\ensuremath{\nabla_{\mathbf{y}}\log p(\mathbf{y})\approx S_{\bm{\theta}}(\mathbf{y};0)}$.
Then, \textbf{Algorithm 2} is executed to obtain the estimated noise
level. Finally, we plug the estimated noise level and the score function
into (\ref{eq:Tweedie}) to compute the estimated channel. We shall
analyze the complexity in Section \ref{subsec:Complexity-Analysis},
which is extremely low since no costly matrix inversion is required.
\floatplacement{algorithm}{t}
\begin{algorithm} 
\caption{Bayes-optimal score-based channel estimator for fully-digital HMIMO systems}
\begin{algorithmic}[1]

\STATE {\bf Input:} Received pilots $\mathbf{y}$, trained DAE parameters $\bm{\theta}$, size of the sliding window $\sqrt{d}\times\sqrt{d}$
\STATE {\bf Output:} Estimated near-field HMIMO channel ${\hat{\mathbf{h}}_\text{MMSE}}$\\
\STATE {\bf Initialize:} $\nabla_\mathbf{y} \log p(\cdot) \gets S_{\bm{\theta}}(\cdot;0)$
\STATE Utilize \textbf{Algorithm 2} to estimate the noise level $\hat{\sigma}_\mathbf{n}$ \\
\STATE Compute the estimated channel $\mathbf{\hat{h}}_{\text{MMSE}}$ by using (\ref{eq:Tweedie})
\STATE {\bf return} $\mathbf{\hat{h}}_{\text{MMSE}}$  \\
\end{algorithmic}
\end{algorithm}

In particular, it is worth noting that the algorithm does not require
any kind of the prior knowledge of the channel or the noise statistics.
Only the received pilot signals $\mathbf{y}$ are required in both
the training and the inference stages. Furthermore, only the DAE network
$\bm{\theta}$ is parameterized in this algorithm, whose training
is unsupervised. Hence, when the EM environment is dynamic, one can
keep updating the parameters $\bm{\theta}$ in an online manner to
track the changes so as to adapt to the new channel distribution,
which will be discussed later in Section \ref{subsec:Online-Adaptation-in}. 

\subsection{Score-Based Estimator in Hybrid Analog-Digital Systems\label{subsec:Extension-to-Hybrid}}

In this section, we shift the focus to the more general hybrid analog-digital
HMIMO systems. We first introduce the general design principle based
on iterative message passing algorithms and explain the practical
difficulties in applying such methods. Then, we discuss how the score-based
estimator and the PCA-based noise level estimation serve as a perfect
match to tackle these challenges and enable Bayes-optimal channel
estimation for HMIMO in unknown EM environments. 

For hybrid analog-digital transceivers, we consider the general system
model (\ref{eq:hybrid-analog-digital-system-model}) for CS-based
channel estimation. Under such circumstances, Bayes-optimal estimation
can be achieved by orthogonal AMP (OAMP)\footnote{The Bayes-optimality of OAMP necessitates the measurement matrix $\mathbf{M}$
to be right-unitarily-invariant \cite{2017Ma}, which covers a large
set of random matrices and can be easily satisfied since $\mathbf{M}$
in the channel estimation depends on the received pilot combiners
which can be manually configured. } \cite{2017Ma,2022Zou}, an iterative algorithm with the following
update rules consisting of a linear estimator (LE) and a non-linear
estimator (NLE), i.e.,

\begin{equation}
\begin{aligned}\begin{array}{cc}
 & \text{OAMP-LE: }\end{array} & \begin{array}{cc}
\mathbf{r}^{(t)}=\mathbf{h}^{(t)}+\mathbf{W}^{(t)}(\mathbf{y}-\mathbf{M}\mathbf{h}^{(t)}),\end{array}\\
\begin{array}{cc}
 & \text{OAMP-NLE: }\end{array} & \begin{array}{cc}
\mathbf{h}^{(t+1)}=\eta_{t}(\mathbf{r}^{(t)}),\end{array}
\end{aligned}
\label{eq:OAMP}
\end{equation}
where the LE is constructed to be de-correlated and the NLE is designed
to be a divergence-free function \cite{2017Ma}. We say that the LE
is de-correlated when $\text{tr}(\mathbf{I}-\mathbf{W}^{(t)}\mathbf{M})=0$
holds, while the divergence-free function can be constructed from
an arbitrary function after subtracting its divergence \cite{2017Ma}.
When both the LE and the NLE are locally MMSE optimal, the replica
Bayes-optimality of the OAMP algorithm was conjectured in \cite{2017Ma}
and rigorously proved in \cite{2020Takeuchi}. The term `orthogonal'
in the name of OAMP derives from the fact that when the conditions
of the LE and the NLE are satisfied, the error after the LE, i.e.,
$\mathbf{e}^{(t)}\triangleq\mathbf{r}^{(t)}-\mathbf{h}$, consists
of independent and identically distributed (i.i.d.) zero-mean Gaussian
elements independent of $\mathbf{h}$, and at the same time, the error
after the NLE, i.e., $\mathbf{q}^{(t+1)}\triangleq\mathbf{h}^{(t+1)}-\mathbf{h}$,
consists of i.i.d. entries independent of $\mathbf{M}$ and $\mathbf{n}$
\cite{2017Ma}. Hence, we have that in the $t$-th iteration, 
\begin{equation}
\mathbf{r}^{(t)}=\mathbf{h}+\mathbf{e}^{(t)},\label{eq:r_t_digital}
\end{equation}
in which $\mathbf{h}$ is the ground-truth HMIMO channel, and $\mathbf{e}^{(t)}\sim\mathcal{N}(\mathbf{0},\sigma_{\mathbf{e}^{(t)}}^{2}\mathbf{I})$
is the effective i.i.d. Gaussian noise vector with $\sigma_{\mathbf{e}^{(t)}}$
denoting the noise level. Hence, the NLE should be the Bayes-optimal
MMSE denoiser to estimate $\mathbf{h}$ from $\mathbf{r}^{(t)}$,
which is essentially the same as the score-based estimator for fully-digital
systems proposed in Section \ref{subsec:Score-Based-Estimator-in-fully-digital}.
In addition, the LE should be the linear MMSE (LMMSE) estimator with
$\mathbf{W}^{(t)}$ being the LMMSE matrix given by
\begin{equation}
\begin{aligned}\mathbf{W}^{(t)} & =\frac{2N}{\text{tr}(\hat{\mathbf{W}}^{(t)}\mathbf{M})}\mathbf{\hat{W}}^{(t)},\text{\,\,where}\\
\mathbf{\hat{W}}^{(t)} & =\sigma_{\mathbf{e}^{(t)}}^{2}\mathbf{M}^{T}(\sigma_{\mathbf{e}^{(t)}}^{2}\mathbf{M}\mathbf{M}^{T}+\sigma_{\mathbf{n}}^{2}\mathbf{I})^{-1}.
\end{aligned}
\label{eq:original-LMMSE-LE}
\end{equation}

The coefficient $\frac{2N}{\text{tr}(\hat{\mathbf{W}}^{(t)}\mathbf{M})}$
is multiplied to ensure that the LE is de-correlated. The final estimated
channel after $t$ iterations is $\mathbf{h}^{(t+1)}$, which, when
converged, is the Bayes-optimal solution. 

\subsubsection{Challenges of OAMP}

Although sounds promising, there are a number of challenges that have
hindered the application of OAMP to near-field HMIMO channel estimation. 

First, the complexity incurred by the LE is prohibitive when the dimension
of the channel $\mathbf{h}\in\mathbb{R}^{2N\times1}$ becomes extremely
large in near-field HMIMO systems, as it requires inverting a $2N\times2N$
general matrix in each iteration, causing a complexity at the order
of $\mathcal{O}(N^{3})$. This is exceedingly large. Second, the noise
statistics is unknown. The level $\sigma_{\mathbf{n}}$ of the Gaussian
noise $\mathbf{n}$ in the EM environment should be accurately estimated
in the first place. Otherwise, both the LE and the NLE cannot be effectively
computed. Third, the optimal MMSE estimator in the NLE cannot be constructed
without knowledge of the prior distribution of the HMIMO channel $\mathbf{h}$,
which, however, is generally not available in practice given the complicated
distribution of the near-field HMIMO channel. One should circumvent
the prior distribution by other techniques so as to implement the
MMSE estimator. In the following, we discuss the practical design
of the LE and the NLE to tackle these challenges mainly on the basis
of the proposed techniques. \begin{algorithm}[!tbp]
\caption{Bayes-optimal score-based channel estimator for hybrid analog-digital HMIMO systems}
\begin{algorithmic}[1]
\STATE {\bf Input:} Measurement matrix $\mathbf{M}$, received pilot signals $\mathbf{y}$, trained DAE parameters $\bm{\theta}$, floor value $\xi$, size of the sliding window $\sqrt{d}\times\sqrt{d}$, error tolerance $\epsilon$, maximum allowed number of iterations $T$ \\
\STATE {\bf Output:} Estimated near-field HMIMO channel ${\hat{\mathbf{h}}_\text{MMSE}}$\\
\STATE {\bf Initialize:} \\ $\mathbf{h}^{(0)} \gets \mathbf{0}, \mathbf{h}^{(1)} \gets \mathbf{M}^\dagger\mathbf{y}, t \gets 0, \nabla_\mathbf{y} \log p(\cdot) \gets S_{\bm{\theta}}(\cdot;0)$ \\
\STATE Obtain $\hat{\sigma}_\mathbf{n}$ by \textbf{Algorithm 2}  \\
\STATE {\bf while} $\|\mathbf{h}^{(t+1)}-\mathbf{h}^{(t)}\| > \epsilon$ {\bf do}
\STATE\hspace{\algorithmicindent} \emph{/*OAMP-LE*/} \\ \hspace{\algorithmicindent} Obtain  $\mathbf{r}^{(t)}$ by   (\ref{eq:OAMP}), (\ref{eq:original-LMMSE-LE}) and (\ref{eq:SVD-LMMSE-LE}) \\ \hspace{\algorithmicindent} Obtain $\hat{\sigma}_{\mathbf{e}^{(t)}}$ by (\ref{eq:effective-noise-level-e})
\STATE\hspace{\algorithmicindent} \emph{/*OAMP-NLE*/} \\ \hspace{\algorithmicindent} Obtain $\mathbf{h}^{(t+1)}$ by (\ref{eq:NLE-MMSE-score})
\STATE\hspace{\algorithmicindent} $t \gets t+1$
\STATE ${\hat{\mathbf{h}}_{\text{MMSE}}} \gets \mathbf{h}^{(t+1)}$
\STATE {\bf return} ${\hat{\mathbf{h}}_{\text{MMSE}}}$  
\end{algorithmic}
\end{algorithm}

\subsubsection{Designing the LE}

The LMMSE matrix in the LE, i.e., (\ref{eq:original-LMMSE-LE}), requires
the computation of a general matrix inverse with prohibitive complexity.
For a general measurement matrix $\mathbf{M}$, the matrix inverse
is essential but could be greatly simplified by using the singular
value decomposition (SVD) of $\mathbf{M}$, i.e., $\mathbf{\mathbf{M=\mathbf{V}\mathbf{D}}\mathbf{U}^{T}}$,
where both $\mathbf{V}$ and $\mathbf{U}$ are orthogonal matrices,
and $\mathbf{D}$ is a diagonal matrix. Plugging the SVD of $\mathbf{M}$
into (\ref{eq:original-LMMSE-LE}), we can get the following results
after some algebra, i.e., 
\begin{equation}
\begin{aligned}\text{\ensuremath{\mathbf{\hat{W}}^{(t)}}} & =\sigma_{\mathbf{e}^{(t)}}^{2}\mathbf{M}^{T}(v_{t}^{2}\mathbf{M}\mathbf{M}^{T}+\sigma_{\mathbf{n}}^{2}\mathbf{I})^{-1}\\
 & =\sigma_{\mathbf{e}^{(t)}}^{2}\mathbf{U}\mathbf{D}^{T}\mathbf{V}^{T}(\sigma_{\mathbf{e}^{(t)}}^{2}\mathbf{\mathbf{V}\mathbf{D}}\mathbf{D}^{T}\mathbf{V}^{T}+\sigma_{\mathbf{n}}^{2}\mathbf{V}\mathbf{I}\mathbf{V}^{T})^{-1}\\
 & =\sigma_{\mathbf{e}^{(t)}}^{2}\mathbf{U}\mathbf{D}^{T}\mathbf{V}^{T}\mathbf{V}(\sigma_{\mathbf{e}^{(t)}}^{2}\mathbf{D}\mathbf{D}^{T}+\sigma_{\mathbf{n}}^{2}\mathbf{I})^{-1}\mathbf{V}^{T}\\
 & =\sigma_{\mathbf{e}^{(t)}}^{2}\mathbf{M}^{T}\mathbf{V}(\sigma_{\mathbf{e}^{(t)}}^{2}\mathbf{D}\mathbf{D}^{T}+\sigma_{\mathbf{n}}^{2}\mathbf{I})^{-1}\mathbf{V}^{T}.
\end{aligned}
\label{eq:SVD-LMMSE-LE}
\end{equation}
Thanks to the SVD operation, we no longer need to calculate a general
matrix inverse like (\ref{eq:original-LMMSE-LE}) in every iteration
$t$. Instead, the new expression requires computing the SVD of matrix
$\mathbf{M}$ only once\footnote{The SVD of $\mathbf{M}$ can be pre-computed and cached.},
and the computation of (\ref{eq:SVD-LMMSE-LE}) becomes much simpler,
since $(\sigma_{\mathbf{e}^{(t)}}^{2}\mathbf{D}\mathbf{D}^{T}+\sigma_{\mathbf{n}}^{2}\mathbf{I})$
is diagonal. The complexity of the inverse is cut down drastically
from $\mathcal{O}(N^{3})$ to $\mathcal{O}(N)$. 

Having solved the complexity issue, another problem is that both the
environmental Gaussian noise level of $\mathbf{n}$, i.e., $\sigma_{\mathbf{n}}$,
and the effective Gaussian noise level of $\mathbf{e}^{(t)}$, i.e.,
$\sigma_{\mathbf{e}^{(t)}}$, are both unknown in practice, but should
be frequently used in OAMP. For $\hat{\sigma}_{\mathbf{n}}$, it can
be estimated from $\mathbf{y}=\text{\ensuremath{\mathbf{Mh}}}+\mathbf{n}$
after treating $\mathbf{Mh}$ as a whole, by using the PCA-based estimator
proposed in \textbf{Algorithm 2}. For $\hat{\sigma}_{\mathbf{e}^{(t)}}$,
we propose two possible methods to estimate it. First, since $\mathbf{r}^{(t)}=\mathbf{h}+\mathbf{e}^{(t)}$
is similar to the fully-digital system model, i.e., (\ref{eq:fully-digital-system-model}),
we could apply the PCA-based noise level estimator to $\mathbf{r}^{(t)}$
to get an estimate of $\hat{\sigma}_{\mathbf{e}^{(t)}}$ in each iteration.
However, this method requires running \textbf{Algorithm 2} once per
iteration, and incurs a high computational complexity. The second
possible option is to utilize the following estimator proposed in
\cite{2013Vila} for $\hat{\sigma}_{\mathbf{e}^{(t)}}$, given by
\begin{equation}
\hat{\sigma}_{\mathbf{e}^{(t)}}^{2}=\text{max}\{\frac{\|\mathbf{y}-\mathbf{M}\mathbf{h}^{(t)}\|^{2}-2KN_{\text{RF}}\hat{\sigma}_{\mathbf{n}}^{2}}{\text{tr}(\mathbf{M}^{T}\mathbf{M})},\xi\},\label{eq:effective-noise-level-e}
\end{equation}
where the function $\text{max}\{\cdot\}$ returns the maximum value
of the inputs, and $\xi>0$ is set as a small positive value to circumvent
numerical instability. The complexity of this approach is much smaller
than the first one since it only involves a simple matrix product,
and the denominator $\text{tr}(\mathbf{M}^{T}\mathbf{M})$ can be
pre-computed. Empirically, we observe that these two methods offer
a similar performance. Hence, we pick the latter due to its relative
lower computational complexity. 

\subsubsection{Designing the NLE}

Based on (\ref{eq:r_t_digital}), $\mathbf{r}^{(t)}$ in each iteration
is the ground-truth channel corrupted by white Gaussian noise $\mathbf{e}^{(t)}\sim\mathcal{N}(\mathbf{0},\sigma_{\mathbf{e}^{(t)}}^{2}\mathbf{I})$,
which is similar to the formulation in fully-digital systems, i.e.,
(\ref{eq:fully-digital-system-model}). After treating $\mathbf{r}^{(t)}$
as $\mathbf{y}$ and $\hat{\sigma}_{\mathbf{e}^{(t)}}$ as $\hat{\sigma}_{\mathbf{n}}$
in \textbf{Algorithm 2}, the NLE is the same as the score-based estimator
in \ref{subsec:Score-Based-Estimator-in-fully-digital}, given by
\begin{equation}
\begin{aligned}\mathbf{h}^{(t+1)}=\eta_{t}(\mathbf{r}^{(t)}) & \triangleq\mathbf{r}^{(t)}+\hat{\sigma}_{\mathbf{e}^{(t)}}^{2}\nabla_{\mathbf{r}^{(t)}}\log p(\mathbf{r}^{(t)})\\
 & \approx\mathbf{r}^{(t)}+\hat{\sigma}_{\mathbf{e}^{(t)}}^{2}S_{\bm{\theta}}(\mathbf{r}^{(t)};0).
\end{aligned}
\label{eq:NLE-MMSE-score}
\end{equation}
As the effective noise level $\sigma_{\mathbf{e}^{(t)}}$ varies in
different numbers of iterations, we pretrain and store a few sets
of DAE parameters corresponding to different noise levels and then
load different DAE parameters $\bm{\theta}$ according to the estimated
noise level $\hat{\sigma}_{\mathbf{e}^{(t)}}$. The complete algorithm
for the score-based estimator in hybrid analog-digital systems is
summarized in \textbf{Algorithm 4}. 

\subsection{Complexity Analysis\label{subsec:Complexity-Analysis}}

We first discuss the complexity expressions in fully-digital systems
and then extend them to hybrid analog-digital systems. 

For fully-digital HMIMO systems, the inference complexity of the proposed
score-based algorithm consists of two parts: the computation of the
score function $S_{\bm{\theta}}(\mathbf{y};0)$ and the PCA-based
estimation of the noise level $\sigma_{\mathbf{n}}$. The former depends
on the specific neural architecture of $S_{\bm{\theta}}(\cdot,\cdot)$,
and costs a constant complexity, denoted by $p$, once the network
is trained. For the latter, we analyze the complexity of \textbf{Algorithm
2} line by line. The complexity of the virtual subarray decomposition
in line 4 is represented as $\mathcal{O}(sd)$. The computation of
the mean and covariance in lines 5 and 6 incurs costs of $\mathcal{O}(sd)$
and $\mathcal{O}(sd^{2})$, respectively. The eigenvalue decomposition
in line 7 has a complexity of $\mathcal{O}(d^{3})$, while sorting
the eigenvalues requires $\mathcal{O}(d^{2})$. The eigenvalue splitting
and the noise level estimation, from lines 8 to 14, have a complexity
of $\mathcal{O}(d^{2})$. Considering that in HMIMO systems, the approximation
$s\thickapprox N$ holds, the complexity of calculating the variance
term is approximately $\mathcal{O}(Nd^{2}+d^{3})$. Therefore, the
overall complexity for the score-based estimator in fully-digital
HMIMO systems, i.e., \textbf{Algorithm 3}, is roughly $\mathcal{O}(Nd^{2}+d^{3}+p)$,
which scales linearly with the number of antennas $N$. In practice,
the fluctuation of the received SNR may not be frequent. In such a
case, it is not a necessity to estimate the received SNR for every
instance of the received pilot signals $\mathbf{y}$. Therefore, the
actual complexity of the proposed score-based algorithm is within
the range of $\mathcal{O}(p)$ and $\text{\ensuremath{\mathcal{O}}}(Nd^{2}+d^{3}+p)$,
which is extremely efficient given the near-optimal performance\footnote{Most previous works on channel estimation assume that the received
SNR is perfectly known \cite{2022Demir,2023An,2023Damico}. As such,
the complexity reduces to $\mathcal{O}(p)$.}. 

For hybrid analog-digital HMIMO systems, we analyze the per-iteration
complexity of \textbf{Algorithm 4}. We first analyze the LE. Thanks
to the SVD-based method, we no longer need to compute a costly general
matrix inversion in each iteration. If the SVD of matrix $\mathbf{M}$
is pre-computed and cached, only the inversion of a diagonal matrix
is required per iteration, which only incurs a complexity of $\mathcal{O}(N)$.
Hence, the complexity of the LE is only dominated by the matrix-vector
product with complexity $\mathcal{O}(NKN_{\text{RF}})$. The overall
complexity of the LE is $\mathcal{O}(NKN_{\text{RF}}+N)$. For the
NLE, the complexity involves one forward propagation of the neural
network $\bm{\theta}$, which is of a constant complexity $p$. Therefore,
the \textit{per-iteration} complexity of the score-based estimator
in hybrid analog-digital systems is $\mathcal{O}(NKN_{\text{RF}}+N+p)$,
dominated by the matrix-vector product. In addition, before the iterations,
we need to run the PCA-based \textbf{Algorithm 2} \textit{only once}
to estimate the Gaussian noise level in the environment, which costs
$\mathcal{O}(Nd^{2}+d^{3})$. 

\section{Simulation Results\label{sec:Simulation-Results}}

\subsection{Simulation Setup\label{subsec:Simulation-Setups}}

We consider uplink channel estimation in a typical HMIMO system whose
general parameters are listed in Table \ref{tab:Key-Simulation-Parameters}.
Specifically, the antenna spacings are configured to be smaller than
the nominal value of half the carrier wavelength $\lambda_{c}$ to
model the quasi-continuous aperture in HMIMO systems. In addition,
the distance between the BS and the scatterer ring is set within the
near-field region, i.e., $S\ll d_{\text{Rayleigh}}$. 

The following baselines are compared in the simulations. 
\begin{table}[t]
\caption{General parameters of the HMIMO system \label{tab:Key-Simulation-Parameters}}

\centering{}%
\begin{tabular}{l|l}
\hline 
\textbf{Parameter} & \textbf{Value}\tabularnewline
\hline 
Number of antennas & $N=1024$\tabularnewline
Carrier frequency & $f_{c}=16$ GHz\tabularnewline
Carrier wavelength & $\lambda_{c}=1.88\times10^{-2}$ m\tabularnewline
Antenna spacing & $d_{a}=\frac{\lambda_{c}}{6}=3.12\times10^{-3}$ m\tabularnewline
Rayleigh distance & $d_{\text{Rayleigh}}=1.09\times10^{3}$ m\tabularnewline
Distance of the scatterer ring & $S=10$ m\tabularnewline
Radius of the scatterer ring & $R=3$ m\tabularnewline
Direction of the scatterer ring & $\Psi=\frac{\pi}{3}$ rad\tabularnewline
Mean of the scatterer angle & $\mu=\frac{\pi}{4}$ rad\tabularnewline
Concentration of the scatterer angle & $\kappa=0$ rad\tabularnewline
\hline 
\end{tabular}
\end{table}
\begin{table}[t]
\begin{centering}
\caption{Performance of the noise level estimation \label{tab:Performance-of-the-SNR}}
\par\end{centering}
\centering{}%
\begin{tabular}{>{\centering}m{1.75cm}|>{\centering}m{1.1cm}|>{\centering}m{0.8cm}|>{\centering}m{0.8cm}|>{\centering}m{0.8cm}|>{\centering}m{0.8cm}}
\hline 
$\sigma_{\mathbf{\mathbf{\bar{n}}}}$ / (SNR) & Method & Bias & Std & RMSE & Percent error\tabularnewline
\hline 
\multirow{3}{1.75cm}{0.5623 (5 dB)} & Oracle & 0.0002 & 0.0087 & 0.0087 & 1.55\%\tabularnewline
\cline{2-6} \cline{3-6} \cline{4-6} \cline{5-6} \cline{6-6} 
 & \textbf{Proposed} & 0.0044 & 0.0153 & 0.0159 & 2.83\%\tabularnewline
\cline{2-6} \cline{3-6} \cline{4-6} \cline{5-6} \cline{6-6} 
 & Sparsity & 0.0344 & 0.0142 & 0.0372 & 6.62\%\tabularnewline
\hline 
\multirow{3}{1.75cm}{0.1778 (15 dB)} & Oracle & $\text{<}$0.0001 & 0.0027 & 0.0027 & 1.52\%\tabularnewline
\cline{2-6} \cline{3-6} \cline{4-6} \cline{5-6} \cline{6-6} 
 & \textbf{Proposed} & 0.0029 & 0.0043 & 0.0052 & 2.92\%\tabularnewline
\cline{2-6} \cline{3-6} \cline{4-6} \cline{5-6} \cline{6-6} 
 & Sparsity & 0.0209 & 0.0087 & 0.0227 & 12.77\%\tabularnewline
\hline 
\multirow{3}{1.75cm}{0.0562 (25 dB)} & Oracle & $\text{<}$0.0001 & 0.0009 & 0.0009 & 1.60\%\tabularnewline
\cline{2-6} \cline{3-6} \cline{4-6} \cline{5-6} \cline{6-6} 
 & \textbf{Proposed} & 0.0004 & 0.0013 & 0.0014 & 2.49\%\tabularnewline
\cline{2-6} \cline{3-6} \cline{4-6} \cline{5-6} \cline{6-6} 
 & Sparsity & 0.0208 & 0.0125 & 0.0243 & 43.24\%\tabularnewline
\hline 
\end{tabular}
\end{table}

\begin{itemize}
\item \textbf{LS}: Least squares estimation. 
\begin{figure*}[t]
\centering{}\subfloat[]{\centering{}\includegraphics[width=0.33\textwidth]{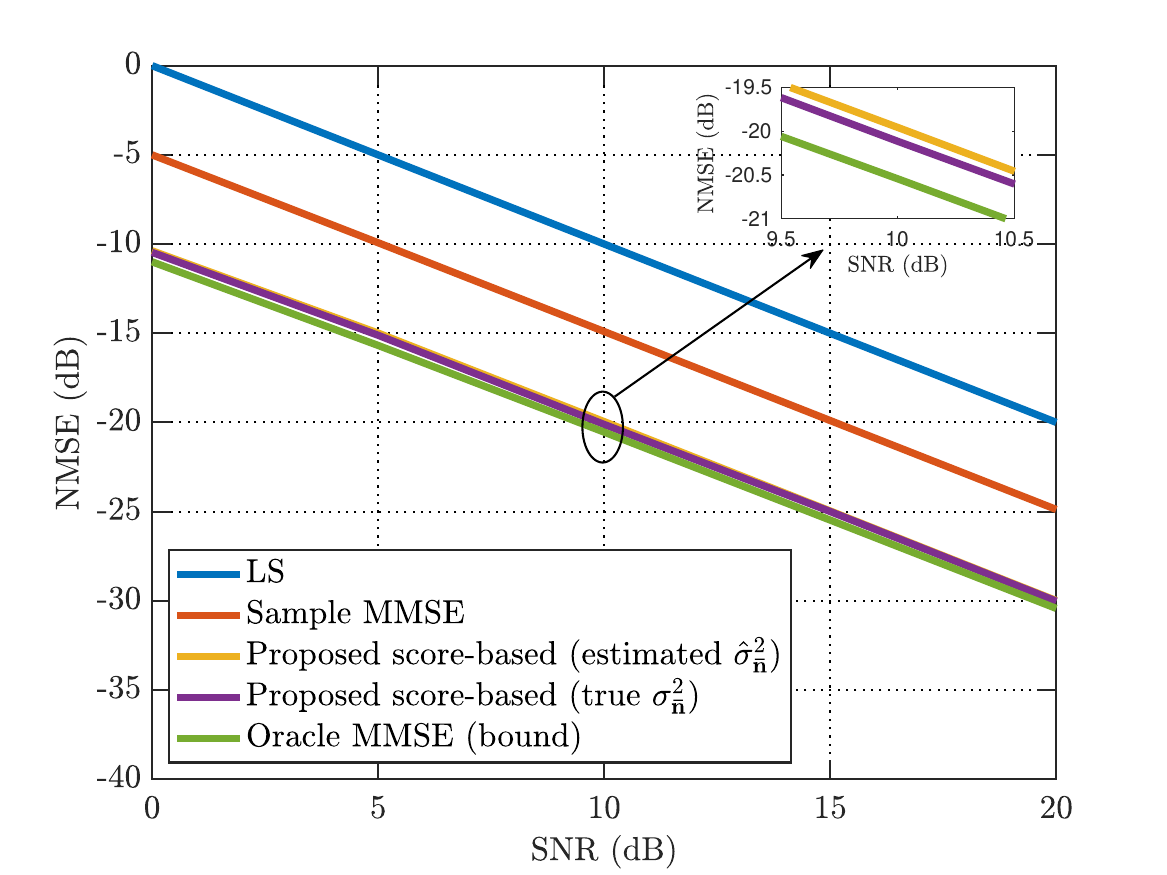}}\subfloat[]{\centering{}\includegraphics[width=0.33\textwidth]{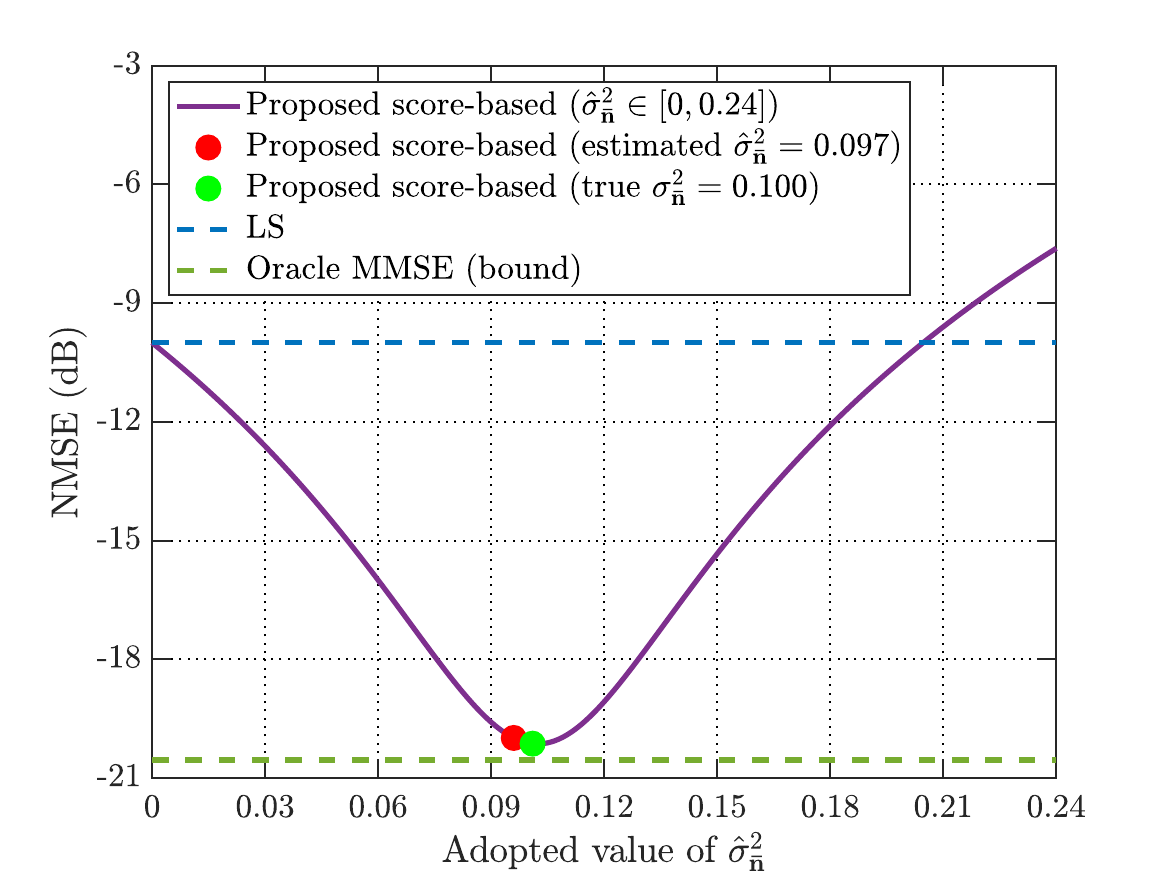}}\subfloat[]{\centering{}\includegraphics[width=0.33\textwidth]{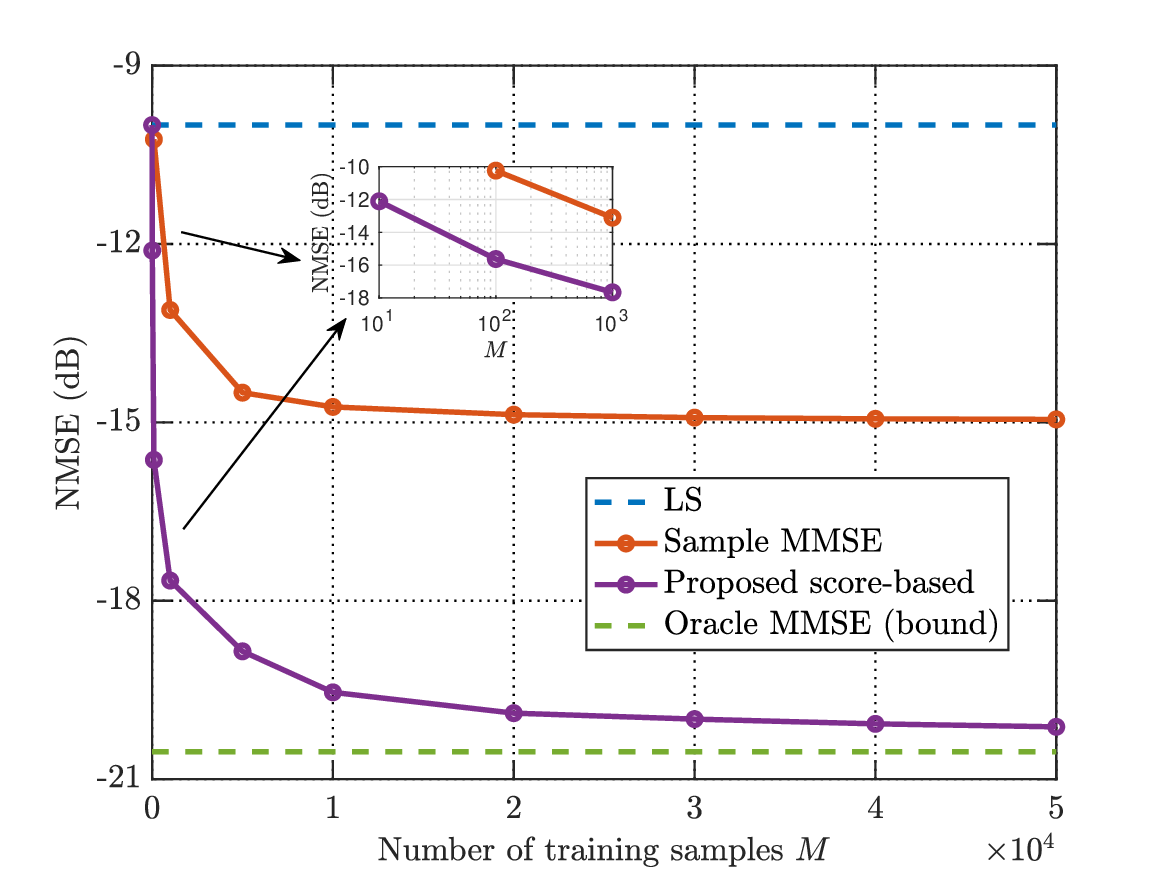}}\caption{Simulation results in fully-digital systems. (a) NMSE as a function
of the received SNR in fully-digital transceivers. (b) The influence
of the accuracy of the received SNR estimation on the NMSE performance,
when the received SNR is 10 dB. (c) Sample complexity of the proposed
score-based estimator, where the NMSE is illustrated as a function
of the number of training samples $M$. The received SNR when plotting
this figure is 10 dB. \label{fig:simulations-fully-digital}}
\end{figure*}
 
\item \textbf{Oracle MMSE}: Oracle MMSE is the MMSE estimator that has a
\textit{perfect knowledge} of both the spatial correlation matrix
$\mathbf{R}_{\text{NF}}$ and the noise level $\sigma_{\bar{\mathbf{n}}}$,
given by 
\begin{equation}
\mathbf{\hat{h}}_{\text{oracle-MMSE}}=\mathbf{R}_{\text{NF}}\text{\ensuremath{\mathbf{M}}}^{H}(\mathbf{M}\mathbf{R}_{\text{NF}}\mathbf{M}^{H}+\sigma_{\mathbf{\bar{n}}}^{2}\mathbf{I})^{-1}\bar{\mathbf{y}}.\label{eq:oracle-MMSE}
\end{equation}
Note that since the HMIMO channel follows a correlated Rayleigh fading,
the linear MMSE is the optimal MMSE estimator that provides the \textit{Bayesian
performance bound}. It is also worth noting that $\mathbf{R}_{\text{RF}}$
is extremely difficult, if not impossible, to be acquired in practice
since it contains $N^{2}$ entries and is thus prohibitive to estimate
when $N$ is particularly large in HMIMO systems. 
\item \textbf{Sample MMSE}: Sample MMSE estimator has the same expression
as (\ref{eq:oracle-MMSE}), but replaces the true spatial correlation
$\mathbf{R}_{\text{NF}}$ with an estimated one $\mathbf{R}_{\text{sample}}$
based upon the testing samples, given by $\mathbf{R}_{\text{sample}}\triangleq\frac{1}{L}\sum_{l=1}^{L}\mathbf{y}_{l}\mathbf{y}_{l}^{H}-\sigma_{\mathbf{\bar{n}}}^{2}\mathbf{I}$
\cite{2023Damico}, in which $L$ is the number of testing samples
and $\mathbf{y}_{l}$ is the $l$-th sample of the received pilots
in the testing dataset. We also utilize the \textit{perfect} noise
level $\sigma_{\mathbf{\bar{n}}}$ in sample MMSE. 
\end{itemize}
During the training stage, the hyper-parameters in the proposed score-based
estimator are selected as $\gamma=0.0004$, $\sigma^{\text{min}}=0.001$,
$\sigma^{\text{max}}=0.1$, $Q=100$, and $M=50\text{,}000$. In addition,
we adopt a batch size of 16 and reduce the learning rate $\gamma$
by half after every 25 epochs. In the inference stage, the shape of
the sliding window in the PCA-based noise level estimation algorithm
is $5\times5$. The performance is evaluated based on the normalized
MSE (NMSE) on a testing dataset with $L=10,000$ samples, i.e., $\text{NMSE}\triangleq\text{\ensuremath{\mathbb{E}}}\nicefrac{\|\mathbf{h}-\mathbf{\hat{h}}\|^{2}}{\|\mathbf{h}\|^{2}}$.
For the neural architecture of the DAE $S_{\bm{\theta}}(\cdot;\cdot)$,
we adopt a simplified UNet architecture \cite{2015Ronneberger}. Note
that depending on the complexity budget, many other prevailing neural
architectures could also be applied \cite{2023Yu-AI}. The SNR definition
in this article refers to the received SNR if not stated otherwise. 

In the following, we first evaluate the accuracy of the PCA-based
noise level estimation, which is a crucial component of the proposed
score-based estimator. Subsequently, we provide extensive simulations
for the score-based estimator in both the fully-digital and the hybrid
analog-digital HMIMO systems, in terms of estimation accuracy, sample
complexity, robustness, and online adaptation capability. 

\subsection{Accuracy of the PCA-Based Noise Level Estimation}

The accuracy of the noise level estimation has an impact on the NMSE
performance of the proposed score-based estimator. Hence, different
from most previous works that have assumed a perfect knowledge of
the noise level $\sigma_{\bar{\mathbf{n}}}$, we alleviate such an
assumption and propose a PCA-based method to estimate it. In Table
\ref{tab:Performance-of-the-SNR}, we list the estimation accuracy
under different SNRs, where $\hat{\sigma}_{\mathbf{\bar{n}}}$ denotes
the estimated noise level. The bias, the standard deviation (std),
and the root MSE (RMSE) of the estimator are given by $\mathbb{E}[|\sigma_{\mathbf{\bar{n}}}-\mathbb{E}[\hat{\sigma}_{\mathbf{\bar{n}}}]|]$,
$\sqrt{\mathbb{E}[(\hat{\sigma}_{\mathbf{\bar{n}}}-\mathbb{E}[\hat{\sigma}_{\mathbf{\bar{n}}}])^{2}]}$,
and $\sqrt{\mathbb{E}[(\sigma_{\bar{\mathbf{n}}}-\hat{\sigma}_{\mathbf{\bar{n}}})^{2}]}$,
respectively. The RMSE gives an overall assessment of the performance,
while the bias and the std reflect the accuracy and robustness of
different estimators. We also provide the percent error, defined as
$\nicefrac{\text{RMSE}}{\sigma_{\mathbf{\bar{n}}}}$, to show the
error as a percentage of the true noise level. We compare the proposed
PCA-based method with the following benchmarks. 
\begin{figure*}[t]
\centering{}\subfloat[]{\centering{}\includegraphics[height=4.5cm]{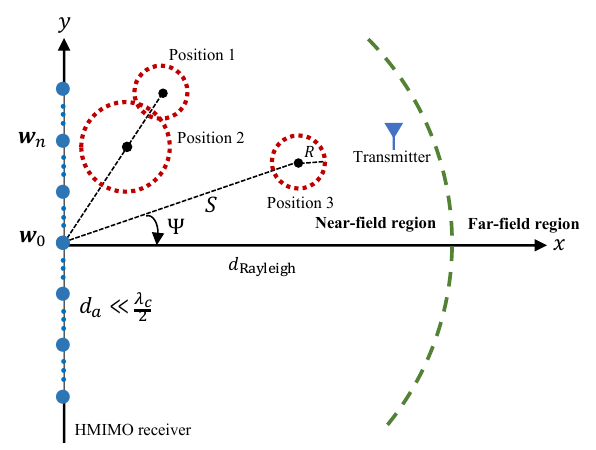}}\subfloat[]{\centering{}\includegraphics[height=4cm]{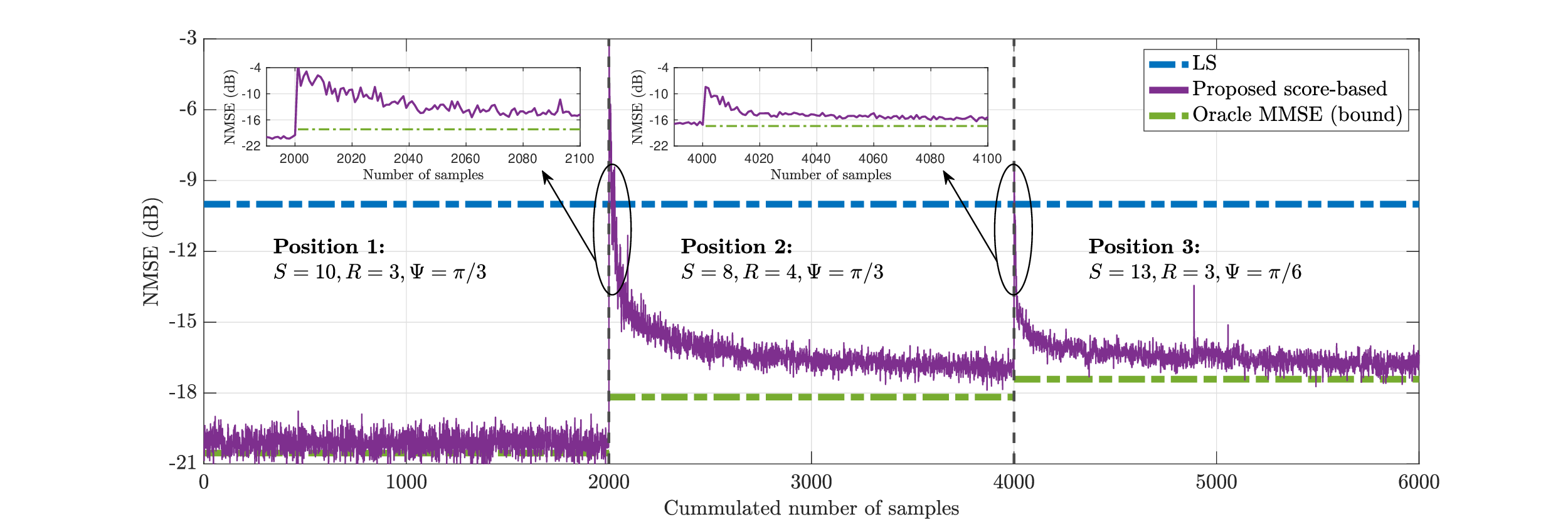}}\caption{Online adaptation in dynamic environments. (a) The positions of the
scatterer rings with different values of direction $\Psi$, radius
$R$, and distance $S$. (b) NMSE versus the number of pilot transmission
slots during the online adaptation process, when the received SNR
is 10 dB. \label{fig:Online-adaptation-dynamic}}
\end{figure*}

\begin{itemize}
\item \textbf{Sparsity-based}: The squared median absolute deviation (MAD)
algorithm for the noise level estimation of sparse signals \cite{2023Gallyas-Sanhueza}.
Similar to \cite{2023Gallyas-Sanhueza}, we utilize the estimator
after transforming the channel into the angular domain. 
\item \textbf{Oracle bound}: Assume perfect knowledge of the ground-truth
channel $\bar{\mathbf{h}}$, and estimate the noise level directly
from $\mathbf{\bar{n}}=\mathbf{\bar{y}}-\mathbf{\bar{h}}$ by using
sample standard deviation, which is the minimum variance unbiased
estimator. 
\end{itemize}
Since the sparsity-based algorithm was originally proposed for the
complex-valued case, we perform the comparison using the complex-valued
version of the proposed PCA-based method, which offers a similar performance
as the real-valued version. 

Table \ref{tab:Performance-of-the-SNR} compares the performance of
different noise level estimators. As shown in the table, the proposed
method has a comparable RMSE with the oracle bound and significantly
outperforms the sparsity-based algorithm. The advantages of our proposal
are consistent across different SNR levels, with a negligible percent
error of less than 3\%. This indicates that the proposed method can
effectively exploit the low-rank property of the near-field spatial
correlation. By contrast, the sparsity-based algorithm performs poorly
since the near-field HMIMO channel lacks sparsity in the angular domain.
Later, in Section \ref{subsec:Robustness-to-SNR}, we will illustrate
that the proposed score-based channel estimator is robust to SNR estimation
errors, and the accuracy of the PCA-based estimator will cause almost
no performance loss compared to using the true noise level. 

\subsection{Estimation Accuracy in Fully-Digital Systems}

In Fig. \ref{fig:simulations-fully-digital}(a), we plot the NMSE
performance as a function of the received SNR $\rho$ in fully-digital
HMIMO transceivers, in which $N_{\text{RF}}=N$. We depict two lines
for the proposed score-based estimator. The yellow one refers to the
performance when the PCA-estimated noise level is adopted in (\ref{eq:Tweedie}),
while the purple line corresponds to the case in which the perfect
noise level is utilized. It is observed that these two lines largely
overlap, indicating that the PCA-estimated noise level provides almost
the same performance as the perfect noise level. Additionally, the
proposed score-based algorithm significantly outperforms the LS and
the sample MMSE estimators by over 10 dB and 5 dB, respectively, in
terms of NMSE, and achieves arguably the same NMSE performance (with
a negligible 0.6 dB drop) as the oracle MMSE performance bound. The
competitive results are consistent in different SNR ranges. Notice
that the oracle MMSE method utilizes the true covariance and noise
level, but the proposed method requires neither. This demonstrates
the strong capability of the unsupervised score-based estimator to
achieve a near-Bayes-optimal performance without any oracle information
in unknown EM environments. 

\subsection{Robustness to Noise Level Estimation Errors\label{subsec:Robustness-to-SNR}}

In Fig. \ref{fig:simulations-fully-digital}(b), we illustrate the
how the accuracy of the estimated noise level will affect the performance
of the proposed score-based estimator. In the simulations, the true
noise level is set as $\sigma_{\mathbf{\bar{n}}}=0.1$, corresponding
to an SNR of 10 dB. We vary the adopted values of $\hat{\sigma}_{\mathbf{\bar{n}}}$
in (\ref{eq:Tweedie}) within $\hat{\sigma}_{\mathbf{\bar{n}}}\in[0,0.24]$,
and plot the corresponding NMSE performance curve in blue. Particularly,
we utilize red and green dots to denote the NMSE achieved with the
estimated and true SNR values, respectively. The LS and the oracle
MMSE algorithms are presented as the performance upper and lower bounds.
It is observed that even when an inexact received SNR is adopted,
the performance of the score-based algorithm is still quite robust
and significantly outperforms the LS method. Also, the estimated received
SNR by the PCA-based method is accurate enough to offer a near-optimal
performance, even in unknown EM environments. 
\begin{figure*}[t]
\centering{}\subfloat[]{\centering{}\includegraphics[width=0.33\textwidth]{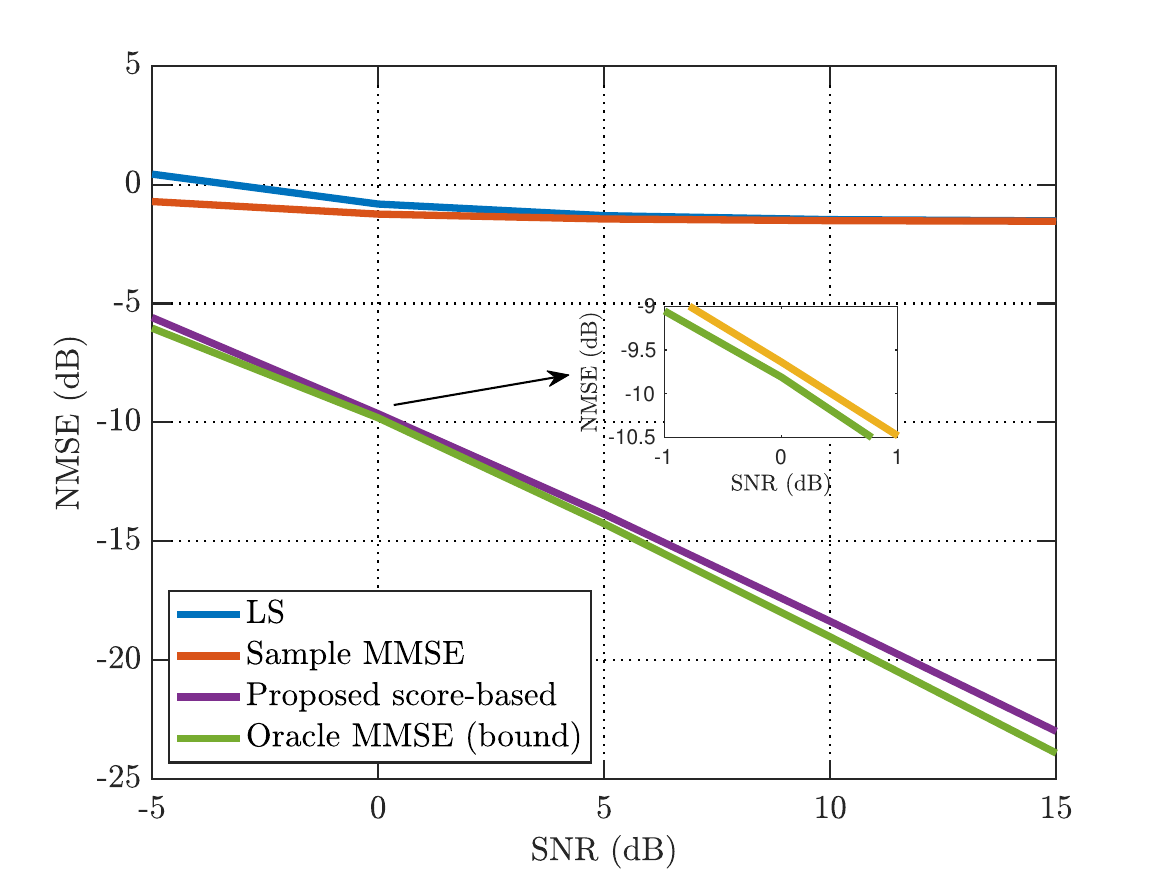}}\subfloat[]{\centering{}\includegraphics[width=0.33\textwidth]{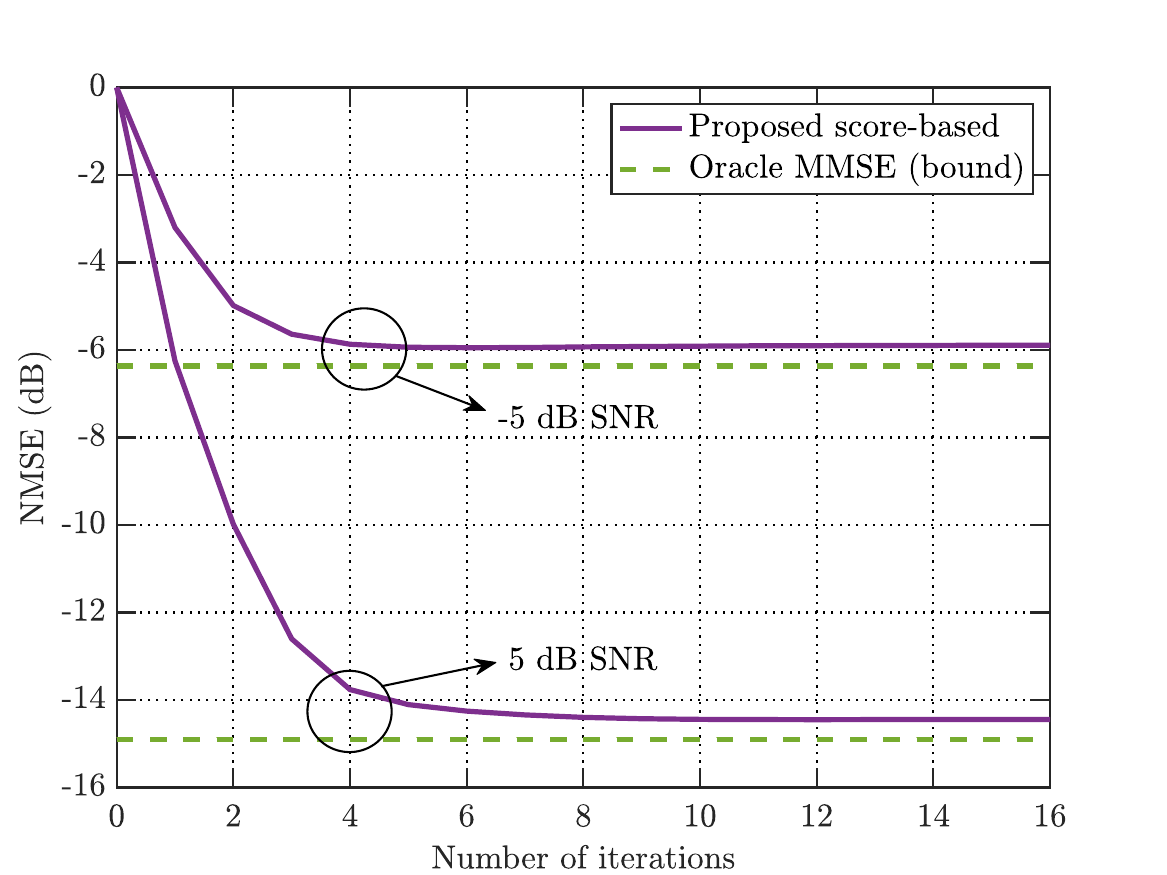}}\subfloat[]{\centering{}\includegraphics[width=0.33\textwidth]{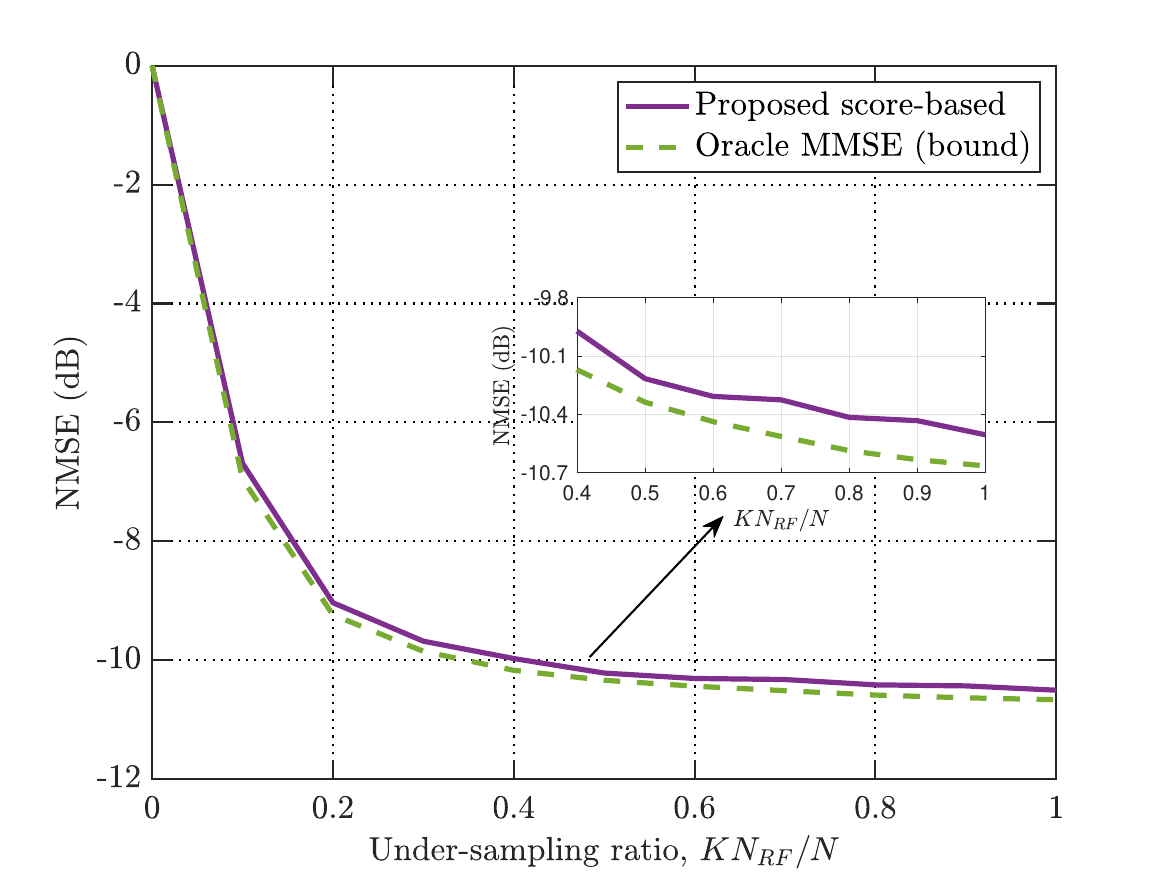}}\caption{Simulation results in hybrid analog-digital systems. (a) NMSE versus
the received SNR when the under-sampling ratio is $KN_{\text{RF}}/N=0.3$.
(b) NMSE as a function of the iteration number when the under-sampling
ratio is $KN_{\text{RF}}/N=0.5$. (c) NMSE as a function of different
under-sampling ratio $KN_{\text{RF}}/N$ when the received SNR is
set as 0 dB. \label{fig:simulations-hybrid-analog-digital}}
\end{figure*}

\subsection{Sample Complexity}

In Fig. \ref{fig:simulations-fully-digital}(c), we study the sample
complexity of the proposed score-based estimator, where the NMSE is
shown as a function of the number of training samples $M$. The received
SNR when plotting the figure is 10 dB. We notice that the gap between
the proposed score-based estimator and the oracle MMSE bound gradually
vanishes when $M$ increases. The NMSE gap shrinks to less than 1
dB at $M=10,000$, and further reduces to less than 0.5 dB when $M=30,000$.
The sample complexity curve of the proposed estimator follows a similar
trend as the sample MMSE method, whose performance both starts to
saturate as $M$ reaches 10,000. Additionally, we are also interested
in the low-sample region when fewer than 1,000 training samples are
available. The region is zoomed in and shown in the subfigure. It
is observed that, the score-based estimator can bring a 2 dB margin
with only $M=10$ training samples, and further bring a 6 dB advantage
with $M=100$ samples, as compared to the LS estimator. This confirms
that the effectiveness of the proposal in different sample regions.
As a final comment, we emphasize again that the training process is
totally unsupervised and only the received pilot signals $\mathbf{y}$
are required. 

\subsection{Online Adaptation in Dynamic Environments\label{subsec:Online-Adaptation-in}}

Since the proposed score-based method is unsupervised, and only requires
the received pilot signals $\mathbf{y}$ during training, it is possible
to update the network parameters $\bm{\theta}$ in an online manner
to adapt to the dynamic EM environments. To illustrate the idea, we
consider a dynamic environment as shown in Fig. \ref{fig:Online-adaptation-dynamic}(a),
which labels three possible positions of the scatterers. We assume
that the scatterers gradually move from positions 1 to 3 to model
the dynamic EM environment, and that each position lasts for 2,000
pilot transmission slots. 

We keep running \textbf{Algorithm 1} so as to update the parameters
of the network parameters $\bm{\theta}$ of the score function. Since
the received pilot signals $\mathbf{y}$ arrive one by one in an online
manner, we utilize a batch size of 1 and update the network parameters
once a new sample is received. In addition, we also change the hyper-parameters
to $\gamma=0.0006$, $\varsigma^{\text{min}}=0.001$, $\varsigma^{\text{max}}=0.1$,
and keep the learning rate $\gamma$ constant. 

In Fig. \ref{fig:Online-adaptation-dynamic}(a), we illustrate the
3 positions of the scatterer ring with different values of direction
$\Psi$, radius $R$, and distance $S$, which are labeled in Fig.
\ref{fig:Online-adaptation-dynamic}(b). We depict the online adaptation
performance as a function of the number of pilot transmission slots.
We assume that the parameters of the network, i.e., $\bm{\theta}$,
are well trained offline for position 1 as described in Section \ref{subsec:Simulation-Setups}.
By contrast, the scatterer rings in positions 2 and 3 are never seen
before in the offline training stage, but suddenly appear in online
deployment. For clarity, we zoom in the transition phase between different
scatterer positions. 

From the results in position 1, it is observed that the online adaptation
will not have negative influence on the performance of the well-trained
network parameters $\bm{\theta}$. During the transition phase from
positions 1 to 2, we first observe an abrupt increase in NMSE. However,
after a few pilot transmission slots, the online adaptation algorithm
quickly recovers the performance to a remarkable level that is close
to the oracle MMSE bound. The performance improves steadily with more
samples. A quite similar curve also appears during the transition
from positions 2 to 3. These results confirm the effectiveness of
the proposed score-based estimator in handling the uncertainties in
the EM environments, which is crucial in practical deployment. The
running time complexity of the online adaptation process is less than
0.01 s per sample on a Nvidia A40 GPU. This ensures a real-time adaptation
to dynamic environments. 

\subsection{Estimation Accuracy in Hybrid Analog-Digital Systems}

In Fig. \ref{fig:simulations-hybrid-analog-digital}(a), we plot the
NMSE as a function of the received SNR when the under-sampling ratio
$KN_{\text{RF}}/N$ equals 0.3 in hybrid analog-digital HMIMO transceivers.
We find that under such a low under-sampling ratio, the performance
gain of the sample MMSE algorithm is marginal as compared to the LS
estimator. By contrast, the performance of the proposed score-based
method still remains close to the oracle MMSE bound across different
SNR levels. As observed, the gap to the oracle bound becomes slighter
larger in the high SNR region. Nevertheless, it is still significantly
better than other baselines. 

In Fig. \ref{fig:simulations-hybrid-analog-digital}(b), we illustrate
the NMSE performance at different number of iterations when the under-sampling
ratio $KN_{\text{RF}}/N$ is 0.5 in hybrid analog-digital systems.
As observed from the figure, the score-based OAMP algorithm converges
rapidly to the oracle bound within only about 5 iterations, which
proves its supreme efficiency and validates our convergence analysis.
It is also interesting to notice that the NMSE is monotonically decreasing
with more iterations. This enables a flexible trade-off between complexity
and performance. 

In Fig. \ref{fig:simulations-hybrid-analog-digital}(c), we examine
the effectiveness of the score-based OAMP algorithm under different
under-sampling ratios when the received SNR is set as 0 dB. The performance
trend of the our proposal aligns well with the oracle bound. The gap
vis-a-vis the bound is consistently smaller than 0.5 dB even under
extremely small under-sampling ratios, e.g., $KN_{\text{RF}}/N=0.1$.
This indicates that our proposed method can seamlessly generalize
to the different pilot lengths $K$ that may be encountered in practice.
It is also interesting to observe that, the NMSE of both curves decreases
quickly when the under-sampling ratio increases from 0 to 0.4, and
gradually saturates thereafter. This confirms that the pilot overhead
for near-field HMIMO channel estimation can be significantly reduced
with proper prior information about the channel. The similar NMSE
trend to the oracle bound also indicates that the score-based method
leads to a near-optimal pilot overhead, and confirms that it effectively
learns the prior distribution of the near-field HMIMO channel in a
totally unsupervised manner. 
\begin{table}[t]
\begin{centering}
\textcolor{black}{\caption{Comparison of the runtime complexity}
}%
\begin{tabular}{c|>{\centering}p{2.2cm}|>{\centering}p{2.2cm}}
\hline 
\textcolor{black}{System architecture} & \textcolor{black}{Oracle \& sample MMSE} & \textcolor{black}{Proposed score-based}\tabularnewline
\hline 
\textcolor{black}{Fully-digital} & \textcolor{black}{290 ms} & \textcolor{black}{2.5 ms}\tabularnewline
\hline 
\textcolor{black}{Hybrid analog-digital} & \textcolor{black}{116 ms} & \textcolor{black}{17.2 ms}\tabularnewline
\hline 
\end{tabular}
\par\end{centering}
\end{table}

\subsection{\textcolor{black}{Runtime Complexity}}

\textcolor{black}{In Section \ref{subsec:Complexity-Analysis}, we
have analyzed the complexity of the proposed algorithms. To offer
a more straightforward comparison, we further provide the runtime
complexity. We provide the runtime complexity as a reference. The
system settings are the same as listed in Table \ref{tab:Key-Simulation-Parameters}.
The under-sampling ratio is set as $KN_{\text{RF}}/N=0.7$ for hybrid
analog-digital systems. The proposed score-based algorithm is set
to terminate after 5 iterations in hybrid analog-digital systems. }

\textcolor{black}{In the fully digital systems, both the oracle MMSE
and the sample MMSE methods necessitate the computation of a high-dimensional
($2N\times2N$) matrix inversion, whose complexity is in the order
of $\mathcal{O}(N^{3})$. The runtime is therefore exceedingly high
in HMIMO systems. By contrast, as analyzed in Section \ref{subsec:Complexity-Analysis},
the proposed score-based method eliminates the matrix inversion and
exhibits a linear complexity with respect to the number of antennas.
Hence, the runtime is significantly shorter. }

\textcolor{black}{In the hybrid analog-digital systems, high-dimensional
matrix inverse is still necessary for the MMSE methods, but the dimension
is decreased to $2KN_{\text{RF}}\times2KN_{\text{RF}}$ ($KN_{\text{RF}}<N$),
leading to a complexity of $\mathcal{O}(K^{3}N_{\text{RF}}^{3})$.
It is still quite large, though shorter than that in fully-digital
systems. By contrast, the complexity of the proposed score-based estimator
is only dominated by the matrix-vector product as analyzed in Section
\ref{subsec:Complexity-Analysis}, which is much shorter compared
to the MMSE methods. The simulation results also support our analysis. }

\subsection{\textcolor{black}{Performance under the Mutual Coupling Effect}}

\textcolor{black}{The mutual coupling (MC) effect will become stronger
with the densely placed antenna elements in HMIMO systems. Hence,
we provide additional experiments to illustrative that effectiveness
of our proposed methods under the MC effect. The MC matrix $\mathbf{Z}\in\mathbb{C}^{N\times N}$
is defined as \cite{2012Stutzman}
\begin{equation}
\mathbf{Z}=Z_{0}(\mathbf{I}+\mathbf{S})(\mathbf{I}-\mathbf{S})^{-1},
\end{equation}
where $Z_{0}$ is the reference antenna impedance, and $\mathbf{S}\in\mathbb{C}^{N\times N}$
denotes the S-parameter matrix of an antenna array, calculated based
on the specific array geometry by using EM simulation tools. Specifically,
we resort to the Antenna Toolbox in Matlab. We refer the readership
to \cite{1983Gupta} for the detailed calculation process. In the
simulations, we utilize the same system parameters described in Table
\ref{tab:Key-Simulation-Parameters}, and define the HMIMO as a dipole
ULA with an element length of $\nicefrac{\lambda_{c}}{2}$ and width
of $\nicefrac{\lambda_{c}}{150}$. We assume that $Z_{0}=50\Omega$,
which is equal to the real-valued port generator resistance \cite{2014Gustafsson}.
Under the MC effect, the effective channel is
\begin{equation}
\mathbf{\bar{h}}_{\text{eff}}=\mathbf{Z}\bar{\mathbf{h}},
\end{equation}
with $\mathbf{\bar{h}}$ being the near-field HMIMO channel as defined
in (\ref{eq:NF-channel}). The effective NF channel should follow
$\mathbf{\bar{h}}_{\text{eff}}\sim\mathcal{CN}(\mathbf{0},\mathbf{R}_{\text{NF-eff}})$,
where $\mathbf{R}_{\text{NF-eff}}\triangleq\mathbf{Z}\mathbf{R}_{\text{NF}}\mathbf{Z}^{H}$
is the effective channel covariance. Similarly, we also transform
$\mathbf{\bar{h}}_{\text{eff}}$ into its real-valued equivalent $\mathbf{h}_{\text{eff}}\triangleq[\Re(\mathbf{\bar{h}}_{\text{eff}})^{T},\Im(\mathbf{\bar{h}}_{\text{eff}})^{T}]^{T}\in\mathbb{R}^{2N\times1}$.
The goal becomes the estimation of the effective channel $\mathbf{h}_{\text{eff}}$,
instead of the original channel $\mathbf{h}$. However, it is worth
noting that the system model and the mathematical problem behind both
remain unchanged. Therefore, the proposed score-based algorithms can
be directly applied to estimate the effective channel under the MC
effect. }

\begin{figure}[t]
\begin{centering}
\textcolor{black}{\includegraphics[width=0.33\textwidth]{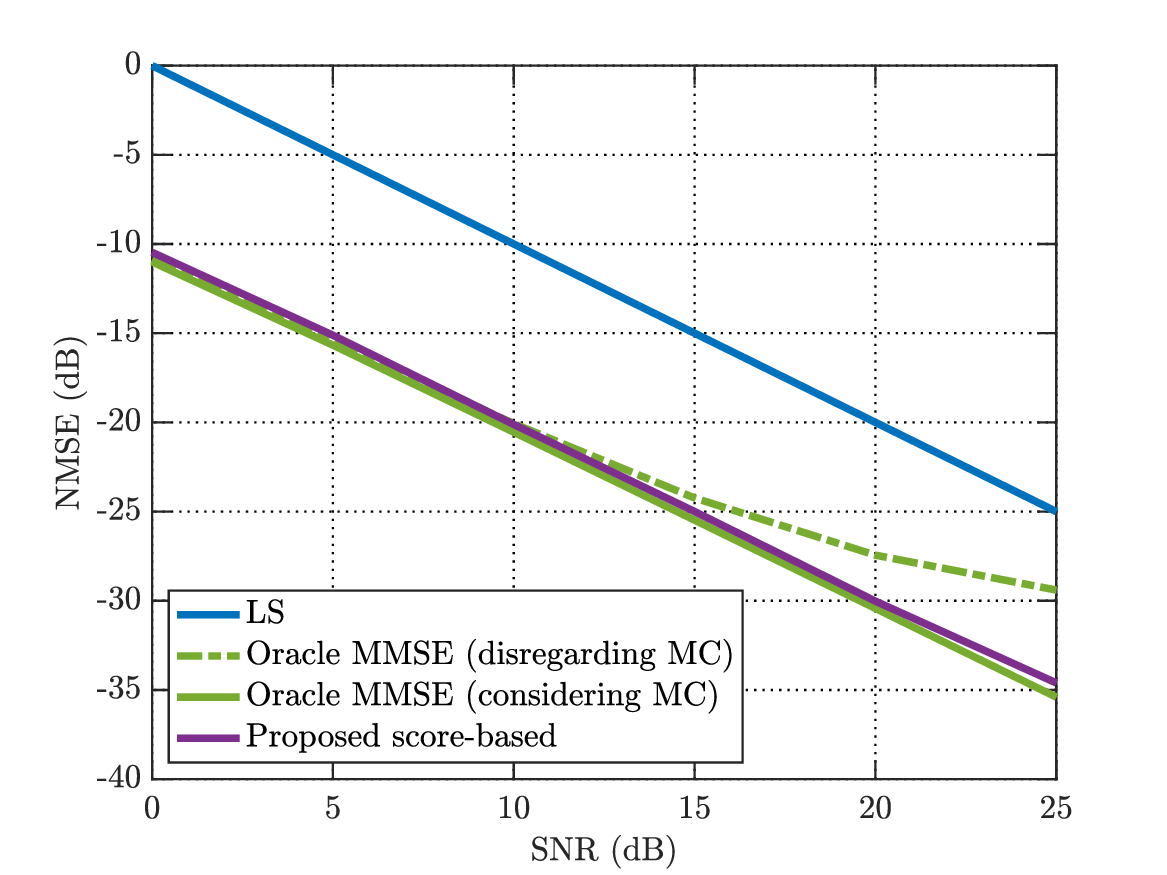}\caption{\textcolor{black}{NMSE as a function of the received SNR $\rho$ in
fully-digital HMIMO transceivers }\textit{\textcolor{black}{under
the MC effect}}\textcolor{black}{. \label{fig:mutual-coupling}}}
}
\par\end{centering}
\end{figure}

\textcolor{black}{In Fig. \ref{fig:mutual-coupling}, we depict the
NMSE performance as a function of the received SNR $\rho$ in fully-digital
HMIMO transceivers under the MC effect}\footnote{\textcolor{black}{The effect of MC on antenna efficiency does not
appear in Fig. \ref{fig:mutual-coupling} because we are analyzing
the received SNR instead of the transmit SNR.}}\textcolor{black}{. Different from the previous simulations, we re-generate
datasets using the }\textit{\textcolor{black}{effective channel }}\textcolor{black}{that
takes MC into account, and train the proposed score-based methods
using the received pilot signals. We compare our proposal with two
oracle MMSE benchmarks. One disregards the MC effect and utilizes
the original NF spatial correlation $\mathbf{R}_{\text{NF}}$ in (\ref{eq:oracle-MMSE}).
By contrast, the second one considers the MC effect and utilizes the
effective covariance $\mathbf{R}_{\text{NF-eff}}$ instead of $\mathbf{R}_{\text{NF}}$.
As observed, these two bounds are close in the low and the mid SNR
regions, while the performance loss of disregarding the MC effect
becomes more apparent in the high SNR scenario. In addition, it is
observed that the proposed score-based methods can still offer the
near Bayes-optimal performance, similar to the oracle MMSE bound considering
the MC effect. This is because the score function estimator is trained
in a unsupervised manner using the received pilot signals $\mathbf{y}$,
which naturally incorporate the distortion caused by the MC effect.
}

\section{Conclusions and Future Directions\label{sec:Conclusion-and-Future}}

Leveraging the connection between the MMSE estimator and the score-function,
in this paper, we proposed an innovative unsupervised learning framework
for channel estimation in near-field HMIMO systems. Different from
existing methods, the proposed algorithms are trained solely based
on the received pilots, without requiring any kind of prior knowledge
of the HMIMO channel or the noise statistics. This enables it to work
in arbitrary and unknown EM environments that may appear in real-world
deployment. The key components of the proposed framework, i.e., learning
the score function, the PCA-based noise level estimation, and the
iterative message passing algorithms, are presented along with theoretical
underpinnings. Extensive simulation results have demonstrated that
our proposal can achieve promising performance in both fully-digital
and hybrid analog-digital systems, reaching an NMSE value close to
the oracle MMSE bound with significant reduction in complexity. Furthermore,
we provided relevant simulation results to validate its strong robustness
and the online adaptation to dynamic EM environments without any priors
or supervision. The results have clearly proven the effectiveness
of the score-based unsupervised learning framework for designing Bayes-optimal
channel estimators in near-field HMIMO systems. As future directions,
it is interesting to extend our proposal to low-resolution HMIMO transceivers
\cite{2018He-lowres} and incorporate advanced meta-learning algorithms
to improve the efficiency of the online adaptation process. 

\appendices{}

\section{Proof of Theorem \ref{thm:The-redundant-eigenvalues}\label{sec:Proof-of-Theorem-Gaussian}}

Define the covariance of $\{\mathbf{y}_{t}\in\mathbb{R}^{2d\times1}\}_{t=1}^{s}$
as 
\begin{equation}
\bm{\Sigma}_{\mathbf{y}_{t}}=\frac{1}{s}\sum_{t=1}^{s}(\mathbf{y}_{t}-\bm{\mu})(\mathbf{y}_{t}-\bm{\mu})^{T},
\end{equation}
in which $\bm{\mu}\triangleq\frac{1}{s}\sum_{t=1}^{s}\mathbf{y}_{t}$
is the mean vector. The eigenvalue decomposition of $\bm{\Sigma}_{\mathbf{y}_{t}}$
is given by
\begin{equation}
\bm{\Sigma}_{\mathbf{y}_{t}}=\mathbf{R}\bm{\text{\ensuremath{\Lambda}}}\mathbf{R}^{T}=\mathbf{R}\text{\text{diag}(\ensuremath{\lambda_{1},\lambda_{2},\ldots,\lambda_{r}})}\mathbf{R}^{T},\,\,\text{where\,\,}r=4d^{2}.
\end{equation}
We can represent $\mathbf{y}_{t}$ by using the eigenvector matrix
$\mathbf{R}$ as a basis, i.e., $\mathbf{y}_{t}=\mathbf{R}\bm{\beta}_{t}$,
with $\mathbf{R}\mathbf{R}^{T}=\mathbf{I}$. Assuming that $\mathbf{h}_{t}$
lies in a $m$-dim subspace ($m\ll r$) and could be represented as
$\mathbf{h}_{t}=\mathbf{A}\bm{\alpha}_{t}$, where $\mathbf{R}=[\mathbf{A},\mathbf{U}]$
with $\mathbf{A}\in\mathbb{R}^{r\times m}$ consisting of the $m$
eigenvectors corresponding to the $m$ largest eigenvalues, we have
that 
\begin{equation}
\mathbf{y}_{t}=\mathbf{R}\bm{\beta}_{t}=[\mathbf{A},\mathbf{U}]\left[\begin{array}{c}
\bm{\alpha}_{t}\\
\mathbf{0}
\end{array}\right]+\mathbf{n}_{t},
\end{equation}
Multiplying both sides with $\mathbf{R}^{T}$, we obtain that 
\begin{equation}
\mathbf{R}^{T}\mathbf{y}_{t}=\left[\begin{array}{c}
\bm{\alpha}_{t}\\
\mathbf{0}
\end{array}\right]+[\mathbf{A},\mathbf{U}]^{T}\mathbf{n}_{t}=\left[\begin{array}{c}
\bm{\alpha}_{t}+\mathbf{A}^{T}\mathbf{n}_{t}\\
\mathbf{U}^{T}\mathbf{n}_{t}
\end{array}\right],
\end{equation}
in which $\mathbf{U}^{T}\mathbf{n}_{t}$ follows $\mathcal{N}(\mathbf{0},\sigma_{\mathbf{n}}^{2}\mathbf{I})$
according to the property of Gaussian distribution. It is observed
that the noise $\mathbf{n}_{t}$ has been separated from the measurement
$\mathbf{y}_{t}$ by matrix $\mathbf{R}$. Hence, the covariance of
$\mathbf{R}^{T}\mathbf{y}_{t}$ is 
\begin{equation}
\frac{1}{s}\sum_{t=1}^{s}(\mathbf{R}^{T}\mathbf{y}_{t})(\mathbf{R}^{T}\mathbf{y}_{t})^{T}=\mathbf{R}^{T}\bm{\Sigma}_{\mathbf{y}_{t}}\mathbf{R}=\bm{\Lambda},
\end{equation}
from which we easily conclude that the redundant eigenvalues follow
a Gaussian distribution with variance $\sigma_{\mathbf{n}}^{2}$. 

\bibliographystyle{IEEEtran}
\bibliography{references_ideas}

\end{document}